%% file: main.tex
\pdfoutput=1
\documentclass[12pt,a4paper]{article}
\usepackage{ifthen} % for conditional statements
\usepackage{rotating}
\usepackage{dashrule}
\usepackage{array} 
\usepackage{dcolumn}
\usepackage{comment}
\usepackage{xcolor}

\usepackage{graphpap} 
\usepackage{makecell} 
\usepackage{rotating} 
\usepackage{tikz}
\usetikzlibrary{patterns.meta}
\usepackage{xcolor}
\usepackage{longtable} % only for template; not usually to be used in PAPERs

\usepackage{lscape}
\usepackage{multirow} 
\usepackage{mathrsfs}
\usepackage{mathtools} %% dcases 
\usetikzlibrary{patterns}
\usepackage{nicefrac} %% 1/2, ...
\usepackage{transparent}
\usepackage{afterpage} 
\usepackage[normalem]{ulem}
\usepackage{cancel}

\newboolean{pdflatex}
\setboolean{pdflatex}{true} % False for eps figures 

\newboolean{articletitles}
\setboolean{articletitles}{true} % False removes titles in references

\newboolean{uprightparticles}
\setboolean{uprightparticles}{true} %True for upright particle symbols

\newenvironment{Tabular}[2][1]
  {\def\arraystretch{#1}\tabular{#2}}
  {\endtabular}

\def\paperauthors{LHCb collaboration} % Leave as is for PAPER and CONF
\def\paperasciititle{Study of Bc+ decays to charmonia plus multihadron final states} % Set ASCII title here
\def\papertitle{Study of 
\decay{\Bc}{\Pchi_{\cquark}\pip} decays} % Latex formatted title

\def\paperkeywords{{High Energy Physics}, {LHCb}} % Comma separated list
\def\papercopyright{\the\year\ CERN for the benefit of the LHCb collaboration} % new since 9/Apr/2018
\def\paperlicence{CC BY 4.0 licence}
\def\paperlicenceurl{https://creativecommons.org/licenses/by/4.0/}

\usepackage{wasysym}

\DeclareMathOperator*{\bigplus}{\scalerel*{+}{\sum}}
\usepackage{scalerel}

\makeatletter
\g@addto@macro\bfseries{\boldmath}
\makeatother

\input{preamble}
\usepackage{longtable} % only for template; not usually to be used in PAPERs

\newcolumntype{d}[1]{D{,}{\,\pm\,}{#1} }
\newcolumntype{f}[1]{D{,}{.}{#1} }

\tikzdeclarepattern{
    name=mylines,
    parameters={
    \pgfkeysvalueof{/pgf/pattern keys/size},
    \pgfkeysvalueof{/pgf/pattern keys/angle},
    \pgfkeysvalueof{/pgf/pattern keys/line width},
    },
bounding box={
    (0,-0.5*\pgfkeysvalueof{/pgf/pattern keys/line width})and
    (\pgfkeysvalueof{/pgf/pattern keys/size},
    0.5*\pgfkeysvalueof{/pgf/pattern keys/line width})},
    tile size={(\pgfkeysvalueof{/pgf/pattern keys/size},
    \pgfkeysvalueof{/pgf/pattern keys/size})},
    tile transformation={rotate=\pgfkeysvalueof{/pgf/pattern keys/angle}},
    defaults={
    size/.initial=5pt,
    angle/.initial=45,
    line width/.initial=.4pt,
    },
code={
    \draw[line width=\pgfkeysvalueof{/pgf/pattern keys/line width}]
    (0,0)-- (\pgfkeysvalueof{/pgf/pattern keys/size},0);
    },
}

\begin{document}

%%%%%%%%%%%%%%%%%%%%%%%%%
%%%%% Title     %%%%%%%%%
%%%%%%%%%%%%%%%%%%%%%%%%%
\renewcommand{\thefootnote}{\fnsymbol{footnote}}
\setcounter{footnote}{1}

% %%%%%%% CHOOSE TITLE PAGE--------
%\onecolumn
\input{title-LHCb-PAPER}

%\twocolumn
% %%%%%%%%%%%%% ---------

\renewcommand{\thefootnote}{\arabic{footnote}}
\setcounter{footnote}{0}

%%%%%%%%%%%%%%%%%%%%%%%%%%%%%%%%
%%%%%  Table of Content   %%%%%%
%%%%%%%%%%%%%%%%%%%%%%%%%%%%%%%%
%%%% Uncomment next 2 lines if desired
%\tableofcontents
%\cleardoublepage

%%%%%%%%%%%%%%%%%%%%%%%%%
%%%%% Main text %%%%%%%%%
%%%%%%%%%%%%%%%%%%%%%%%%%

\pagestyle{plain} % restore page numbers for the main text
\setcounter{page}{1}
\pagenumbering{arabic}

%% Uncomment during review phase. 
%% Comment before a final submission.

% You can include short sections directly in the main tex file.
% However, for larger papers it is desirable to split the text into
% several semiautonomous files, which can be revised independently.
% This is especially useful when developing a document in
% collaboration with several people, since then different parts can be
% edited independently.  This type of file organization is shown here.
% 

\input{body}

%%% temporatily commented out 
\input{acknowledgements}

%% \clearpage
%% \input{supplementary-app}

\usetikzlibrary{patterns}

\clearpage
\addcontentsline{toc}{section}{References}
\bibliographystyle{LHCb}
\bibliography{main,standard,LHCb-PAPER,LHCb-CONF,LHCb-DP,LHCb-TDR}
 
%%% temporaril commneted out 
\clearpage
\input{Authorship_LHCb-PAPER-2023-039}

\end{document}

%% file: preamble.tex
\usepackage[top=1in, bottom=1.25in, left=1in, right=1in]{geometry}

\usepackage{microtype}
\usepackage{lineno}  % for line numbering during review
\usepackage{xspace} % To avoid problems with missing or double spaces after
\usepackage{caption} %these three command get the figure and table captions automatically small

\usepackage{graphicx}  % to include figures (can also use other packages)
\usepackage{color}
\usepackage{colortbl}
\graphicspath{{./figs/}} % Make Latex search fig subdir for figures
\DeclareGraphicsExtensions{.pdf,.PDF,png,.PNG}

\usepackage{amsmath} % Adds a large collection of math symbols
\usepackage{amssymb}
\usepackage{amsfonts}
\usepackage{upgreek} % Adds in support for greek letters in roman typeset

\newcommand*\patchAmsMathEnvironmentForLineno[1]{%
\expandafter\let\csname old#1\expandafter\endcsname\csname #1\endcsname
\expandafter\let\csname oldend#1\expandafter\endcsname\csname
end#1\endcsname
 \renewenvironment{#1}%
   {\linenomath\csname old#1\endcsname}%
   {\csname oldend#1\endcsname\endlinenomath}%
}
\newcommand*\patchBothAmsMathEnvironmentsForLineno[1]{%
  \patchAmsMathEnvironmentForLineno{#1}%
  \patchAmsMathEnvironmentForLineno{#1*}%
}
\AtBeginDocument{%
\patchBothAmsMathEnvironmentsForLineno{equation}%
\patchBothAmsMathEnvironmentsForLineno{align}%
\patchBothAmsMathEnvironmentsForLineno{flalign}%
\patchBothAmsMathEnvironmentsForLineno{alignat}%
\patchBothAmsMathEnvironmentsForLineno{gather}%
\patchBothAmsMathEnvironmentsForLineno{multline}%
\patchBothAmsMathEnvironmentsForLineno{eqnarray}%
}

\usepackage{hyperxmp}

\usepackage[pdftex,
            pdfauthor={\paperauthors},
            pdftitle={\paperasciititle},
            pdfkeywords={\paperkeywords},
            pdfcopyright={Copyright (C) \papercopyright},
            pdflicenseurl={\paperlicenceurl}]{hyperref}

\usepackage[colorinlistoftodos,textsize=scriptsize]{todonotes}

\usepackage[all]{hypcap} % Internal hyperlinks to floats.

\input{lhcb-symbols-def} % Add in the predefined LHCb symbols

\usepackage{cite} % Allows for ranges in citations
\usepackage{mciteplus}

%% file: lhcb-symbols-def.tex
\usepackage{xspace} 
\usepackage{upgreek}

\def\lhcb   {\mbox{LHCb}\xspace}

\def\babar  {\mbox{BaBar}\xspace}
\def\belle  {\mbox{Belle}\xspace}

\def\cdf    {\mbox{CDF}\xspace}

\def\MagUp {\mbox{\em Mag\kern -0.05em Up}\xspace}

\ifthenelse{\boolean{uprightparticles}}%
{
 
 \def\Pgamma      {\ensuremath{\upgamma}\xspace}

 \def\Pmu         {\ensuremath{\upmu}\xspace}

 \def\Ppi         {\ensuremath{\uppi}\xspace}

 \def\Pchi        {\ensuremath{\upchi}\xspace}                 
 \def\Ppsi        {\ensuremath{\uppsi}\xspace}

 \def\PDelta      {\ensuremath{\Delta}\xspace}                 
 \def\PXi         {\ensuremath{\Xi}\xspace}                 
 \def\PLambda     {\ensuremath{\Lambda}\xspace}                 
 \def\PSigma      {\ensuremath{\Sigma}\xspace}                 
 \def\POmega      {\ensuremath{\Omega}\xspace}                 
 \def\PUpsilon    {\ensuremath{\Upsilon}\xspace}
 \let\oldPi\Pi
 \def\PPi         {\ensuremath{\oldPi}\xspace}

 \def\PB      {\ensuremath{\mathrm{B}}\xspace}                 
                  
 \def\PD      {\ensuremath{\mathrm{D}}\xspace}

 \def\PJ      {\ensuremath{\mathrm{J}}\xspace}                 
 \def\PK      {\ensuremath{\mathrm{K}}\xspace}

 \def\PW      {\ensuremath{\mathrm{W}}\xspace}

 \def\Pb      {\ensuremath{\mathrm{b}}\xspace}                 
 \def\Pc      {\ensuremath{\mathrm{c}}\xspace}

 \def\Pi      {\ensuremath{\mathrm{i}}\xspace}

 \def\Pp      {\ensuremath{\mathrm{p}}\xspace}                 
 \def\Pq      {\ensuremath{\mathrm{q}}\xspace}                 
                  
 \def\Ps      {\ensuremath{\mathrm{s}}\xspace}

 \def\thebaroffset{0.0em}
}
{
 
 \def\Pgamma      {\ensuremath{\gamma}\xspace}

 \def\Pmu         {\ensuremath{\mu}\xspace}

 \def\Ppi         {\ensuremath{\pi}\xspace}

 \def\Pchi        {\ensuremath{\chi}\xspace}                 
 \def\Ppsi        {\ensuremath{\psi}\xspace}                 
                  
 \mathchardef\PDelta="7101
 \mathchardef\PXi="7104
 \mathchardef\PLambda="7103
 \mathchardef\PSigma="7106
 \mathchardef\POmega="710A
 \mathchardef\PUpsilon="7107
 \mathchardef\PPi="7105
                  
 \def\PB      {\ensuremath{B}\xspace}                 
                  
 \def\PD      {\ensuremath{D}\xspace}

 \def\PJ      {\ensuremath{J}\xspace}                 
 \def\PK      {\ensuremath{K}\xspace}

 \def\PW      {\ensuremath{W}\xspace}

 \def\Pb      {\ensuremath{b}\xspace}                 
 \def\Pc      {\ensuremath{c}\xspace}

 \def\Pi      {\ensuremath{i}\xspace}

 \def\Pp      {\ensuremath{p}\xspace}                 
 \def\Pq      {\ensuremath{q}\xspace}                 
                  
 \def\Ps      {\ensuremath{s}\xspace}

 \def\thebaroffset{0.18em}
}
\newcommand{\offsetoverline}[2][\thebaroffset]{\kern #1\overline{\kern -#1 #2}}%

\makeatletter
\ifcase \@ptsize \relax% 10pt
  \newcommand{\miniscule}{\@setfontsize\miniscule{4}{5}}% \tiny: 5/6
\or% 11pt
  \newcommand{\miniscule}{\@setfontsize\miniscule{5}{6}}% \tiny: 6/7
\or% 12pt
  \newcommand{\miniscule}{\@setfontsize\miniscule{5}{6}}% \tiny: 6/7
\fi
\makeatother

\DeclareRobustCommand{\optbar}[1]{\shortstack{{\miniscule (\rule[.5ex]{1.25em}{.18mm})}
  \\ [-.7ex] $#1$}}

\def\mumu       {{\ensuremath{\Pmu^+\Pmu^-}}\xspace}

\def\g      {{\ensuremath{\Pgamma}}\xspace}

\def\Wp     {{\ensuremath{\PW^+}}\xspace}

\def\quark     {{\ensuremath{\Pq}}\xspace}
\def\quarkbar  {{\ensuremath{\overline \quark}}\xspace}

\def\squark    {{\ensuremath{\Ps}}\xspace}

\def\cquark    {{\ensuremath{\Pc}}\xspace}
\def\cquarkbar {{\ensuremath{\overline \cquark}}\xspace}

\def\bquark    {{\ensuremath{\Pb}}\xspace}

\def\pion   {{\ensuremath{\Ppi}}\xspace}
\def\piz    {{\ensuremath{\pion^0}}\xspace}
\def\pip    {{\ensuremath{\pion^+}}\xspace}
\def\pim    {{\ensuremath{\pion^-}}\xspace}

\def\kaon    {{\ensuremath{\PK}}\xspace}
\def\KorKbar {\kern \thebaroffset\optbar{\kern -\thebaroffset \PK}{}\xspace}

\def\Kp      {{\ensuremath{\kaon^+}}\xspace}
\def\Km      {{\ensuremath{\kaon^-}}\xspace}

\def\KS      {{\ensuremath{\kaon^0_{\mathrm{S}}}}\xspace}

\def\Kstarp  {{\ensuremath{\kaon^{*+}}}\xspace}

\def\D       {{\ensuremath{\PD}}\xspace}

\def\DorDbar {\kern \thebaroffset\optbar{\kern -\thebaroffset \PD}\xspace}
\def\Dz      {{\ensuremath{\D^0}}\xspace}

\def\Dp      {{\ensuremath{\D^+}}\xspace}
\def\Dm      {{\ensuremath{\D^-}}\xspace}

\def\DpDm    {\ensuremath{\Dp {\kern -0.16em \Dm}}\xspace}

\def\Dstarp  {{\ensuremath{\D^{*+}}}\xspace}

\def\B       {{\ensuremath{\PB}}\xspace}

\def\BorBbar {\kern \thebaroffset\optbar{\kern -\thebaroffset \PB}\xspace}

\def\Bd      {{\ensuremath{\B^0}}\xspace}

\def\BdorBdbar {\kern \thebaroffset\optbar{\kern -\thebaroffset \Bd}\xspace}
\def\Bu      {{\ensuremath{\B^+}}\xspace}

\def\Bp      {{\ensuremath{\Bu}}\xspace}

\def\Bs      {{\ensuremath{\B^0_\squark}}\xspace}

\def\BsorBsbar {\kern \thebaroffset\optbar{\kern -\thebaroffset \Bs}\xspace}
\def\Bc      {{\ensuremath{\B_\cquark^+}}\xspace}

\def\jpsi     {{\ensuremath{{\PJ\mskip -3mu/\mskip -2mu\Ppsi}}}\xspace}
\def\psitwos  {{\ensuremath{\Ppsi{(2{\mathrm{S}})}}}\xspace}

\def\chic     {{\ensuremath{\Pchi_\cquark}}\xspace}
\def\chiczero {{\ensuremath{\Pchi_{\cquark 0}}}\xspace}
\def\chicone  {{\ensuremath{\Pchi_{\cquark 1}}}\xspace}

\def\chictwo  {{\ensuremath{\Pchi_{\cquark 2}}}\xspace}

\def\Y#1S{\ensuremath{\PUpsilon{(#1S)}}\xspace}

\def\proton      {{\ensuremath{\Pp}}\xspace}

\def\Lz          {{\ensuremath{\PLambda}}\xspace}

\def\LorLbar     {\kern \thebaroffset\optbar{\kern -\thebaroffset \PLambda}\xspace}

\def\Lb           {{\ensuremath{\Lz^0_\bquark}}\xspace}

\def\BF         {{\ensuremath{\mathcal{B}}}\xspace}
\def\BR         {\BF}

\newcommand{\decay}[2]{\ensuremath{#1\!\to #2}\xspace} 

\def\to                 {\ensuremath{\rightarrow}\xspace}

\def\AT#1     {\ensuremath{A_{\mathrm{T}}^{#1}}\xspace}           % 2

\def\C#1      {\ensuremath{\mathcal{C}_{#1}}\xspace}                       % 9
\def\Cp#1     {\ensuremath{\mathcal{C}_{#1}^{'}}\xspace}                    % 7
\def\Ceff#1   {\ensuremath{\mathcal{C}_{#1}^{\mathrm{(eff)}}}\xspace}        % 9  
\def\Cpeff#1  {\ensuremath{\mathcal{C}_{#1}^{'\mathrm{(eff)}}}\xspace}       % 7
\def\Ope#1    {\ensuremath{\mathcal{O}_{#1}}\xspace}                       % 2
\def\Opep#1   {\ensuremath{\mathcal{O}_{#1}^{'}}\xspace}                    % 7

\newcommand{\nospaceunit}[1]{\ensuremath{\text{#1}}}       
\newcommand{\aunit}[1]{\ensuremath{\text{\,#1}}}       
\newcommand{\tev}{\aunit{Te\kern -0.1em V}\xspace}
\newcommand{\gev}{\aunit{Ge\kern -0.1em V}\xspace}
\newcommand{\mev}{\aunit{Me\kern -0.1em V}\xspace}
\newcommand{\kev}{\aunit{ke\kern -0.1em V}\xspace}
\newcommand{\ev}{\aunit{e\kern -0.1em V}\xspace}
 
\newcommand{\mevc}{\ensuremath{\aunit{Me\kern -0.1em V\!/}c}\xspace}
\newcommand{\gevc}{\ensuremath{\aunit{Ge\kern -0.1em V\!/}c}\xspace}
\newcommand{\mevcc}{\ensuremath{\aunit{Me\kern -0.1em V\!/}c^2}\xspace}
\newcommand{\gevcc}{\ensuremath{\aunit{Ge\kern -0.1em V\!/}c^2}\xspace}
\def\mum  {\ensuremath{\,\upmu\nospaceunit{m}}\xspace}

\def\fb   {\ensuremath{\aunit{fb}}\xspace}
\def\invfb   {\ensuremath{\fb^{-1}}\xspace}

\def\gsim{{~\raise.15em\hbox{$>$}\kern-.85em
          \lower.35em\hbox{$\sim$}~}\xspace}
\def\lsim{{~\raise.15em\hbox{$<$}\kern-.85em
          \lower.35em\hbox{$\sim$}~}\xspace}

\def\sPlot{\mbox{\em sPlot}\xspace}

\def\pt         {\ensuremath{p_{\mathrm{T}}}\xspace}

\def\evtgen     {\mbox{\textsc{EvtGen}}\xspace}

\def\geant      {\mbox{\textsc{Geant4}}\xspace}

\def\photos     {\mbox{\textsc{Photos}}\xspace}

\def\pythia     {\mbox{\textsc{Pythia}}\xspace}

\def\tell1  {TELL1\xspace}
\def\ukl1   {UKL1\xspace}

\newcommand{\lhcborcid}[1]{\href{https://orcid.org/#1}{\hspace*{0.1em}\raisebox{-0.45ex}{\includegraphics[width=1em]{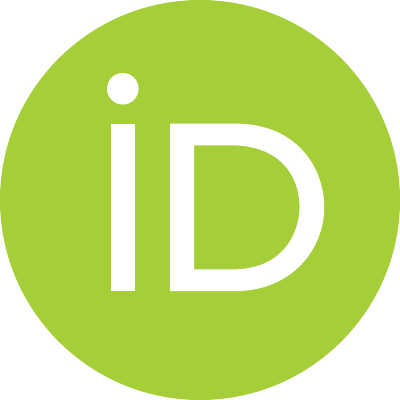}}}}

\def\BcTochicjpi       {\decay{\Bc}{\chic\pip}}
\def\BcTojpsipi       {\decay{\Bc}{\jpsi\pip}}
\def\BcTochiconepi    {\decay{\Bc}{\chicone\pip}}
\def\BcTochictwopi    {\decay{\Bc}{\chictwo\pip}}

%% file: title-LHCb-PAPER.tex
% $Id: title-LHCb-PAPER.tex 122889 2018-08-17 17:59:55Z pkoppenb $
% ===============================================================================
% Purpose: LHCb-PAPER journal paper title page template
% Author: 
% Created on: 2010-09-25
% ===============================================================================

%%%%%%%%%%%%%%%%%%%%%%%%%
%%%%%  TITLE PAGE  %%%%%%
%%%%%%%%%%%%%%%%%%%%%%%%%
\begin{titlepage}
\pagenumbering{roman}

% Header ---------------------------------------------------
\vspace*{-1.5cm}
\centerline{\large EUROPEAN ORGANIZATION FOR NUCLEAR RESEARCH (CERN)}
\vspace*{1.5cm}
\noindent
\begin{tabular*}{\linewidth}{lc@{\extracolsep{\fill}}r@{\extracolsep{0pt}}}
\ifthenelse{\boolean{pdflatex}}% Logo format choice
{\vspace*{-1.5cm}\mbox{\!\!\!\includegraphics[width=.14\textwidth]{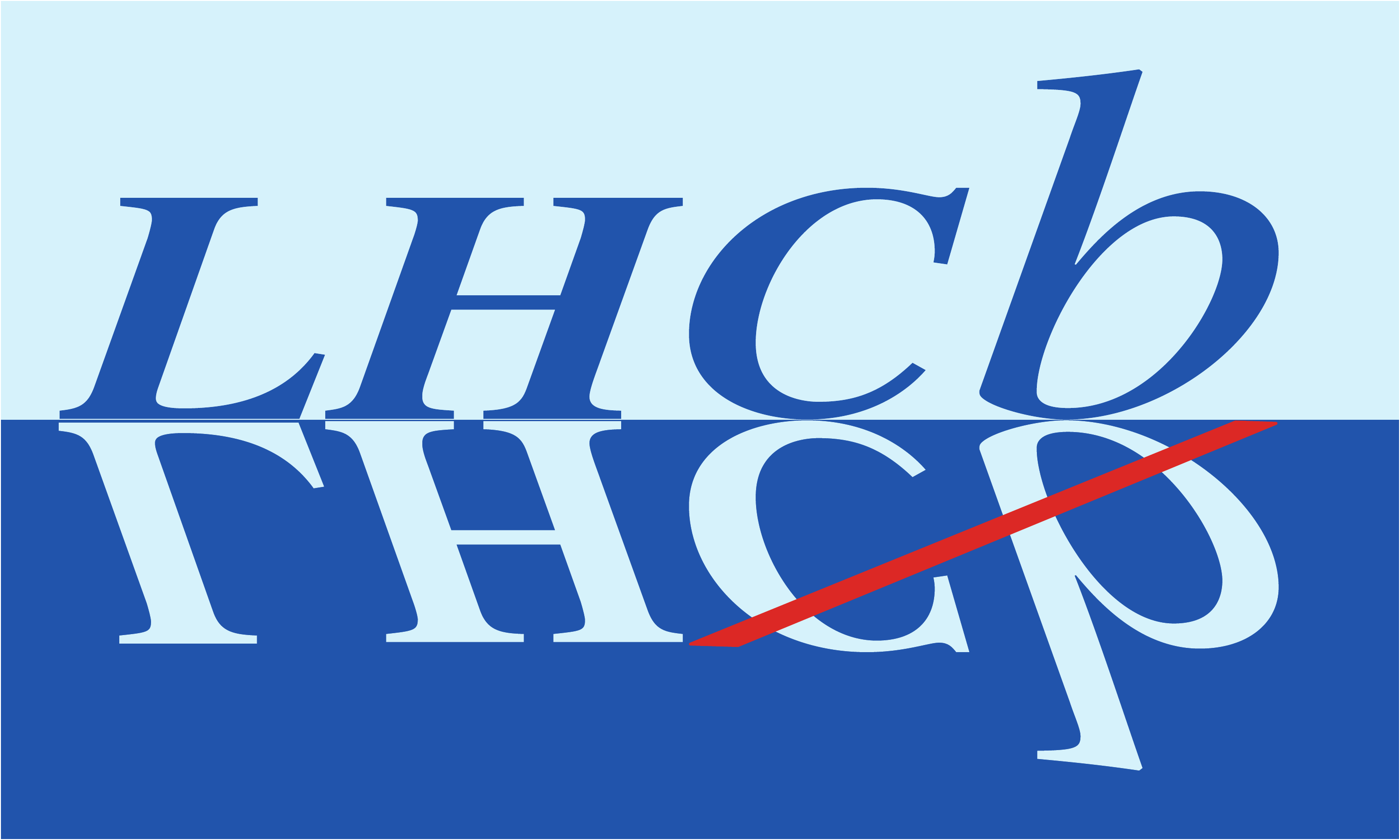}} & &}%
{\vspace*{-1.2cm}\mbox{\!\!\!\includegraphics[width=.12\textwidth]{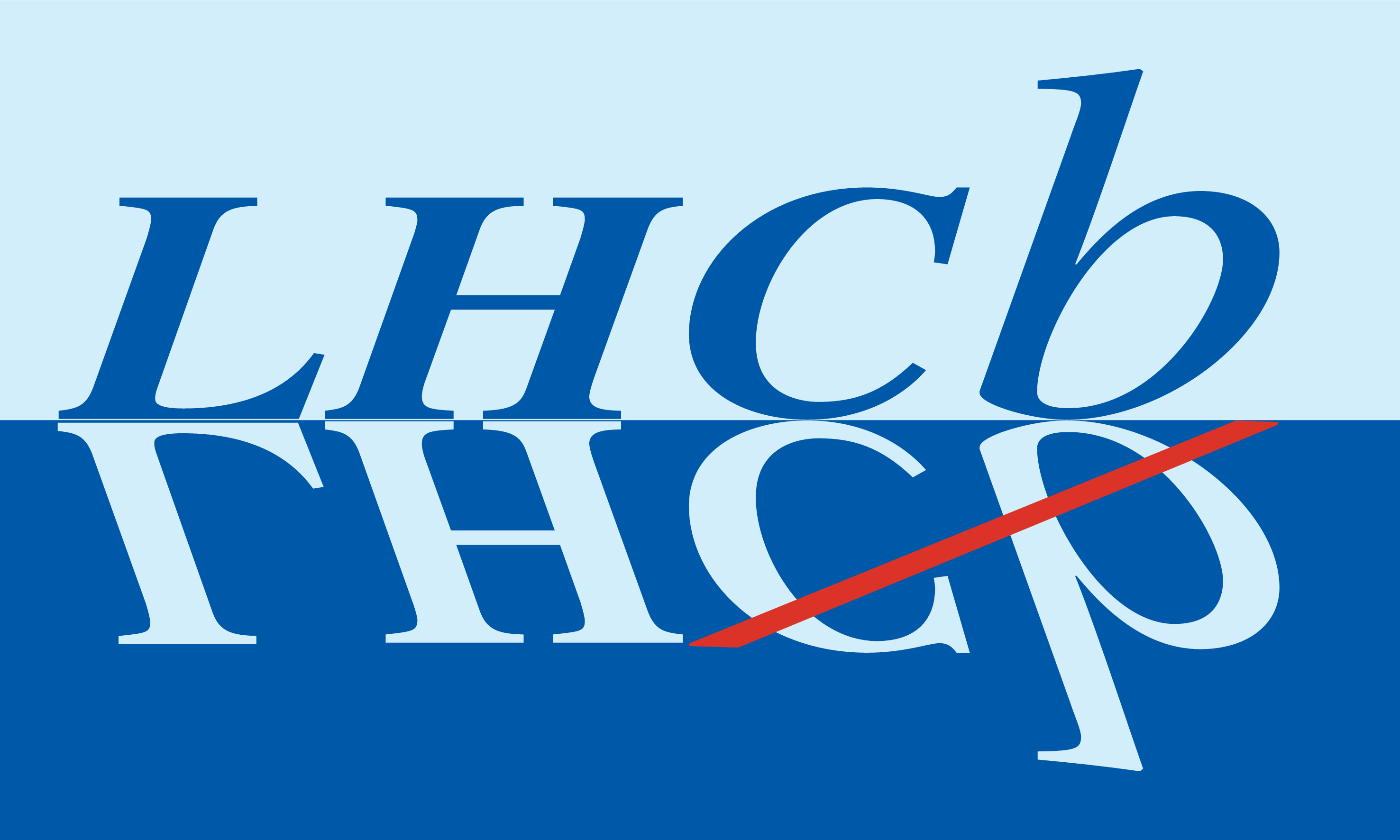}} & &}%
\\
 & & CERN-EP-2023-274 \\  % ID 
 & & LHCb-PAPER-2023-039 \\  % ID 
%% & & \today \\ % Date - Can also hardwire e.g.: 23 March 2010
 &  & December 18, 2023 \\ 
%% & & Month XX, 2023\\ % Date - Can also hardwire e.g.: 23 March 2010
%% & & version 0.10
%% & &
% not in paper \hline
\end{tabular*}

\vspace*{3.0cm}

% Title --------------------------------------------------
{\normalfont\bfseries\boldmath\huge
\begin{center}
% DO NOT EDIT HERE. Instead edit macro in main.tex to keep metadata correct
  \papertitle 
\end{center}
}

\vspace*{0.2cm}

% Authors -------------------------------------------------
\begin{center}
%In the footnote, replace 'paper' by 'Letter' in case of submission to PRL or PLB 
% Edit macro in main.tex to keep metadata correct
\paperauthors\footnote{Authors are listed at the end of this paper.}
\end{center}

%\vspace{\fill}

% Abstract -----------------------------------------------
\begin{abstract}
\noindent 
A~study of \BcTochicjpi~decays is reported 
using proton\nobreakdash-proton collision data, 
collected with the~\lhcb detector at 
centre\nobreakdash-of\nobreakdash-mass energies of 
7, 8, and 13\tev, 
corresponding to an~integrated luminosity of~$9\invfb$.
The~decay \BcTochictwopi is observed for the~first time,
with a~significance exceeding 
seven~standard deviations.
The~relative branching 
fraction with respect to
the~\mbox{\BcTojpsipi}~decay is measured to be
$$
\dfrac{\BR_{\BcTochictwopi}}{\BR_{\BcTojpsipi}} =  
  0.37 \pm 0.06 \pm 0.02 \pm 0.01\,,
$$
where the first uncertainty is statistical,
the~second is systematic, and 
the~third is due to the~knowledge of 
the~\decay{\chictwo}{\jpsi\g}
branching fraction. 
No~significant \mbox{$\decay{\Bc}{\chicone\pip}$}~signal 
is observed 
and an~upper limit for the relative
branching fraction
for the~\BcTochiconepi and \BcTochictwopi decays of 
$$
\dfrac{\BR_{\BcTochiconepi}}{\BR_{\BcTochictwopi}} 
< 0.49
$$
is set at the~90\% confidence level. 
\end{abstract}

%%\vspace*{6.5cm}
\vspace{\fill}

\begin{center}
 %% Submitted  to JHEP 
 Published in \href{https://doi.org/10.1007/JHEP02(2024)173}{JHEP 02 (2024) 173}
 
\end{center}

\vspace*{0.5cm}

{\footnotesize 
% Edit macro in main.tex to keep metadata correct
\centerline{\copyright~\papercopyright. \href{\paperlicenceurl}{\paperlicence}.}}
\vspace*{2mm}

\end{titlepage}

%%%%%%%%%%%%%%%%%%%%%%%%%%%%%%%%
%%%%%  EOD OF TITLE PAGE  %%%%%%
%%%%%%%%%%%%%%%%%%%%%%%%%%%%%%%%

%  empty page follows the title page ----
\newpage
\setcounter{page}{2}
\mbox{~}
%\newpage
%
%% Author List ----------------------------
%%  You need to get a new author list!
%\input{LHCb_authorlist.tex}

\cleardoublepage

%% file: body.tex
\section{Introduction}
\label{sec:Introduction}
%% In the Standard Model of particle physics 
The~\Bc meson,
discovered in 1998 by 
the~\cdf collaboration~\cite{PhysRevLett.81.2432, 
PhysRevD.58.112004}
at the~Tevatron 
%% $\proton\antiproton$
collider, 
%%
%% are unique because they contain 
is the~only known meson that contains
two different heavy\nobreakdash-flavour quarks, 
charm and beauty. 
The~high $\bquark$\nobreakdash-quark production cross\nobreakdash-section at 
the~Large Hadron
Collider\,(LHC)~\cite{LHCb-PAPER-2010-002,
LHCb-PAPER-2011-003,
LHCb-PAPER-2011-043,
LHCb-PAPER-2013-004,
LHCb-PAPER-2013-016,
LHCb-PAPER-2015-037} 
enables the~\lhcb, ATLAS and CMS experiments 
to study in detail the~production, 
decays and other properties 
of the~\Bc meson~\cite{LHCb-PAPER-2011-044, 
LHCb-PAPER-2012-054, 
LHCb-PAPER-2013-010, 
LHCb-PAPER-2013-021, 
LHCb-PAPER-2013-044, 
LHCb-PAPER-2013-047, 
LHCb-PAPER-2014-009, 
LHCb-PAPER-2014-039, 
LHCb-PAPER-2014-050, 
LHCb-PAPER-2014-060, 
CMS:2014oqy,
LHCb-PAPER-2015-024, 
ATLAS:2015jep,
LHCb-PAPER-2016-001, 
LHCb-PAPER-2016-022, 
LHCb-PAPER-2016-020, 
LHCb-PAPER-2016-055, 
LHCb-PAPER-2016-058, 
LHCb-PAPER-2017-035, 
LHCb-PAPER-2017-045, 
CMS:2017ygm,         %% lifetime 
CMS:2019uhm,         %% Bc(2S)
LHCb-PAPER-2019-007, %% Bc(2S)
LHCb-PAPER-2019-033, 
LHCb-PAPER-2020-003, 
LHCb-PAPER-2021-023,
LHCb-PAPER-2021-034,
LHCb-PAPER-2022-025}. 
The~\Bc~meson has 
a~rich set of 
decay modes 
since either of the~heavy quarks can 
decay while the~other
behaves as a~spectator quark, 
or both quarks can annihilate 
via a~virtual \Wp~boson. 
Decays of the~\Bc meson to charmonium and 
light hadrons can be  described
using the~quantum chromodynamics~(QCD) 
factorisation approach~\cite{Bauer:1986bm,
Wirbel:1988ft},
which relies on the~form factors of 
the~\mbox{\decay{\Bc}
{[\cquark\cquarkbar]\Wp}}~transition~\cite{Gershtein:1994jw,
Gershtein:1997qy,
Kiselev:1999sc,
Kiselev:2000pp,
Ebert:2002pp}, 
and on the~universal spectral function 
for the~virtual \Wp~boson fragmenting
into light hadrons~\cite{Likhoded:2009ib, 
Likhoded:2013iua, 
Berezhnoy:2011is}.

The~decays of beauty hadrons into final states with 
 P\nobreakdash-wave charmonium, %%  states, 
such as \chicone and \chictwo mesons,
have been extensively studied by 
the~ARGUS~\cite{ARGUS:1991qht},
CLEO~\mbox{\cite{CLEO:1994mwq,CLEO:2000emb}},
\babar~\mbox{\cite{BaBar:2001kpq,
BaBar:2004htr,
BaBar:2005kmj,
BaBar:2005pcw,
BaBar:2006fjg,
Aubert:2008ae,
BaBar:2019hzd}},
\belle~\mbox{\cite{Belle:2005eoz,
Belle:2006eqz,
Belle:2008qeq,
Belle:2011wdj,
Belle:2015opn,
Belle:2017psv,
Belle:2019avj}},
CDF~\cite{CDF:2002vbg}
and 
LHCb~\mbox{\cite{LHCb-PAPER-2013-024,
LHCb-PAPER-2017-011,
LHCb-PAPER-2018-018,
LHCb-PAPER-2021-003}}
collaborations.
In~\mbox{$\decay{\PB}{\chic\PK^{(*)}}$}~decays\footnote{In 
this paper the~symbol $\chic$ 
denotes the~$\chicone$ and $\chictwo$~states together
and inclusion of charge-conjugate
decays is implied.}
a~significant suppression 
of the~tensor $\chictwo$~state 
relative to the~vector $\chicone$~state 
is observed
in accordance
with expectations from 
QCD factorisation~\cite{Beneke:2008pi}. 
In~contrast,
for decays of the~beauty baryon \Lb, 
the~\mbox{$\decay{\Lb}{\chicone\proton\Km}$} 
and~\mbox{$\decay{\Lb}{\chictwo\proton\Km}$}~decay widths 
are found to be approximately the~same~\cite{LHCb-PAPER-2017-011}.
A~similar trend is observed for 
the~\mbox{$\decay{\Lb}{\chicone\proton\pim}$} 
and~\mbox{$\decay{\Lb}{\chictwo\proton\pim}$}~decays~\cite{LHCb-PAPER-2021-003}.
For~decays of the~\Bc~meson 
the~inverted pattern
%%  of the~decays
%% via the~tensor $\chictwo$ 
%% and vector $\chicone$ states 
is expected 
with 
a~relative suppression of the~decays
with the~$\chicone$~meson in the~final state~\cite{
PhysRevD.65.014017,
PhysRevD.82.034019,
PhysRevD.74.074008,
PhysRevD.73.054024,
vvKiselev_2002,
PhysRevD.97.033001,
Wang_2012,
ZHU2018359,
Chao-Hsi_2001}. 
However, the~experimental information 
on the~decays of the~\Bc~meson
into $\mathrm{P}$\nobreakdash-wave charmonium 
states is very limited. 
Currently no decays of the~\Bc~meson into 
the~$\chicone$ and 
$\chictwo$ states are observed, and there is only 
evidence for the~\mbox{$\decay{\Bc}{\left(\decay{\chiczero}{\Kp\Km}\right)\pip}$}~decay
with the~scalar $\chiczero$~meson in the~final state~\cite{LHCb-PAPER-2016-022}.
%% More experimental information is 
Additional measurements are 
required to test 
the~theory predictions
and, in particular,
to clarify the~role of QCD factorisation
in \Bc~meson decays.

This paper reports a~study 
of~\mbox{\BcTochicjpi} decays
using the~radiative decays
\mbox{$\decay{\chic}{\jpsi\g}$}.
%% of the~$\chic$~mesons.
A~sample of \mbox{$\decay{\Bc}{\jpsi\pip}$}~decays 
is used for the~normalisation
and 
a~high\nobreakdash-yield 
%% low\nobreakdash-background 
sample of ~\mbox{$\decay{\Bu}{\chicone\Kp}$}~decays 
is used to~calibrate 
the~detector resolution and mass scale. 
The~analysis is based 
on proton\nobreakdash-proton\,($\proton\proton$) collision data, 
corresponding to an~integrated 
luminosity of~9\,\invfb,
collected with the~\lhcb detector 
at~centre-of-mass energies of 7, 8, and 13\,\tev.

\section{Detector and simulation}
\label{sec:Detector}

The \lhcb detector~\cite{Alves:2008zz,LHCb-DP-2014-002} is a single-arm forward
spectrometer covering the~pse\-udora\-pi\-dity range \mbox{$2<\eta <5$},
designed for the study of particles containing $\bquark$~or~$\cquark$~quarks. 
The~detector includes a high-precision tracking system consisting of a 
silicon-strip vertex detector surrounding the \proton\proton interaction
region~\cite{LHCb-DP-2014-001}, a large-area silicon-strip detector located
upstream of a dipole magnet with a bending power of about $4{\mathrm{\,Tm}}$,
and three stations of silicon-strip detectors and straw
drift tubes~\cite{LHCb-DP-2013-003,LHCb-DP-2017-001} placed downstream of the magnet. 
The tracking system provides a measurement of the momentum of charged particles
with a relative uncertainty that varies from $0.5\%$ at low momentum to $1.0\%$~at~$200 \gevc$. 
The~momentum scale is calibrated using samples of $\decay{\jpsi}{\mumu}$ 
and $\decay{\Bu}{\jpsi\Kp}$~decays collected concurrently
with the~data sample used for this analysis~\cite{LHCb-PAPER-2012-048,LHCb-PAPER-2013-011}. 
The~relative accuracy of this
procedure is estimated to be $3 \times 10^{-4}$ using samples of other
fully reconstructed $\bquark$~hadrons, 
$\PUpsilon$~and
$\KS$~mesons.
The~minimum distance between a~track and 
a~primary 
$\proton\proton$\nobreakdash-collision vertex\,(PV)\cite{Bowen:2014tca,Dziurda:2115353}, 
the~impact parameter, %% \,(IP), 
is~measured with a~resolution of $(15+29/\pt)\mum$, where \pt is 
the~component 
of the~momentum transverse 
to the~beam, in\,\gevc. Different types of charged hadrons
are distinguished using information from 
two ring\nobreakdash-imaging Cherenkov 
detectors\,(RICH)~\cite{LHCb-DP-2012-003}. Photons,
electrons and hadrons are identified 
by a~calorimeter system consisting of scintillating\nobreakdash-pad 
and preshower detectors, 
an electromagnetic and 
a~hadronic calorimeter. Muons are~identified by a~system 
composed of alternating layers of iron and multiwire proportional chambers~\cite{LHCb-DP-2012-002}.

The online event selection is performed by a trigger~\cite{LHCb-DP-2012-004}, 
which consists of a hardware stage, based on information from the calorimeter and muon systems,
followed by a~software stage, which performs a full event reconstruction. 
The~hardware trigger selects muon candidates 
with high transverse momentum 
or dimuon candidates with a~high value of 
the~product
of the~transverse momenta of the~two muons.
In~the~software trigger, 
two 
oppositely-charged muons are required to form 
a~good\nobreakdash-quality
vertex that is significantly 
displaced from any~PV,
and the~mass of the~$\mumu$~pair 
is required to  
exceed~$2.7\gevcc$.

Simulated 
samples of 
\mbox{$\decay{\Bc}{\chic\pip}$},
%% \mbox{$\decay{\Bc}{\chictwo\pip}$},
\mbox{$\decay{\Bc}{\jpsi\pip}$}
and 
\mbox{$\decay{\Bu}{\chic\Kp}$}~decays 
are used 
to model the~signal mass shapes, 
%% tune 
optimise 
the~selection requirements 
and compute the~efficiencies needed to determine 
the~branching fraction ratios.
The~\pythia~\cite{Sjostrand:2007gs} 
generator with a~specific 
\lhcb configuration~\cite{LHCb-PROC-2010-056}
is used to simulate
\proton\proton collisions with~\Bu~meson production.
For~\Bc~meson production
the~{\sc{BcVegPy}}~generator~\cite{Chang:2003cq,
Chang:2005hq,
Wang:2012ah,
Wu:2013pya}
is used. 
It~is~based on the~full perturbative QCD
%% complete 
calculations at the~lowest 
order $\upalpha_{\mathrm{s}}^4$
via the~dominant 
gluon\nobreakdash-gluon fusion 
process 
\mbox{$\decay{\mathrm{gg}}
{\Bc\left(\B^{*+}_{\cquark}\right)
+ \cquarkbar + \bquark}$}
and 
neglecting contributions from the~quark pair annihilation 
\mbox{$\decay{\mathrm{\quark\quarkbar}}
{\Bc\left(\B^{*+}_{\cquark}\right) + 
\cquarkbar + \bquark}$}~channel\cite{Chang:1992jb,
Chang:1994aw,
Chang:1996jt,
Kolodziej:1995nv,
Berezhnoy:1996ks}.
The~generator is
interfaced with the~\pythia~parton shower and hadronisation model.
Decays of unstable particles are described by 
the~\evtgen 
package~\cite{Lange:2001uf}, 
in which final-state radiation is generated using \photos~\cite{davidson2015photos}. 

The~interaction of the~generated particles 
with the~detector,
and its response, are implemented using
the~\geant 
toolkit~\cite{Allison:2006ve,*Agostinelli:2002hh} 
as described in Ref.~\cite{LHCb-PROC-2011-006}.
The~transverse momentum and rapidity spectra of 
the~\Bc mesons in simulated samples
are adjusted to match those observed in 
a~high\nobreakdash-yield, 
low\nobreakdash-background sample of reconstructed
\BcTojpsipi decays. 
The~detector response used for the~identification of pions and kaons 
is sampled from 
the~\mbox{$\decay{\Dstarp}{\left(\decay{\Dz}{\Km\pip}\right)\pip}$}
and \mbox{$\decay{\KS}{\pip\pim}$} control channels~\cite{LHCb-DP-2012-003,
LHCb-DP-2018-001}. 
To~account for imperfections in the~simulation of
charged particle reconstruction, 
the~track  reconstruction efficiency
determined from simulation 
is corrected using 
\mbox{$\decay{\jpsi}{\mumu}$}~calibration samples~\cite{LHCb-DP-2013-002}.
Samples of 
\mbox{$\decay{\Bp}{\jpsi\left(\decay{\Kstarp}{\Kp\piz}\right)}$}~decays 
are used to
correct the~photon reconstruction 
efficiency in 
simulation~\cite{LHCb-PAPER-2012-022, 
LHCb-PAPER-2012-053, 
Govorkova:2015vqa,
Govorkova:2124605,
Belyaev:2016cri}.

\section{Event selection}
\label{sec:Selection}
 
The signal~\BcTochicjpi
and control \mbox{$\decay{\Bu}{\chic\Kp}$}~candidates 
are reconstructed using 
\mbox{$\decay{\chic}{\jpsi\g}$}
followed by~\mbox{$\decay{\jpsi}{\mumu}$}~decays. 
The~dimuon final state of the~\jpsi~meson is used to reconstruct 
\BcTojpsipi~candidates that are used as a normalisation channel.
A~loose preselection
is applied,
followed by a~multivariate classifier based on a decision
tree with gradient boosting\,({\sc{BDTG}})~\cite{Breiman}.

Muon, pion and kaon candidates are identified 
by combining information from
the~RICH, calorimeter and muon detectors. 
The~candidates are required to have transverse momenta
larger than 550\mevc and 
200\mevc for muon and hadron candidates, respectively.
Pions and kaons are required to have a momentum 
between 3.2 and 150\gevc to ensure good performance 
of particle identification in 
the~RICH detectors~\cite{LHCb-PROC-2011-008,
LHCb-DP-2012-003}. 
To reduce combinatorial
backgrounds from particles produced in \proton\proton~interactions, 
only tracks that are inconsistent with originating from any PV are used.

Pairs of oppositely charged 
muons consistent with originating from a~common vertex 
are combined to form ~\mbox{$\decay{\jpsi}{\mumu}$}~candidates. 
The~reconstructed mass of the~\mumu~pair 
is required to be
in the~range $3.0<m_{\mumu}<3.2\gevcc$. 
The~position of
the~reconstructed dimuon vertex is required to 
be inconsistent with that of any reconstructed~PV.

The~\mbox{$\decay{\chic}{\jpsi\g}$}~candidates
are formed by combining 
the~selected \jpsi~candidates with 
photon candidates
that have been reconstructed using clusters in
the~electromagnetic calorimeter. 
Only~clusters that are not matched to the~trajectory 
of a~track extrapolated from the tracking
system~\cite{Terrier:691743,
AbellanBeteta:2020amj} are used.
The~transverse energy of photon candidates 
is required to exceed~400\mev.
The $\chic$ candidates are selected 
in the~\mbox{$\jpsi\g$}~mass region between
3.4 and 3.7\gevcc, which covers 
both the~\mbox{$\decay{\chicone}{\jpsi\g}$}
and \mbox{$\decay{\chictwo}{\jpsi\g}$}~decays.

%%  To~form the~\Bc~and~\Bu~candidates, 
The~selected $\chic$ and \jpsi~candidates 
 are combined with charged tracks identified as 
 pions or kaons 
 to form \BcTochicjpi,
\mbox{$\decay{\Bu}{\chic\Kp}$}
and \BcTojpsipi candidates. 
A kinematic fit~\cite{Hulsbergen:2005pu}
that constrains 
the~three\nobreakdash-track 
combination 
%% final-state particles 
to form a~common vertex is performed, in which 
the~mass of the~\mumu~combination is
set equal to the~known \jpsi mass~\cite{PDG2023}  
and the~\B~candidate is constrained to originate from 
the~associated~PV.\footnote{Each \Bc or \Bu~candidate 
is associated with the~PV that 
yields the smallest $\chi^2_{\rm{IP}}$, where $\chi^2_{\rm{IP}}$ is defined 
as the~difference in the~vertex\nobreakdash-fit 
$\chi^2$ of a~given PV reconstructed with and
without the~particle under consideration.}
A~good fit quality is required to further suppress
combinatorial background. 
The~measured decay time of the~selected candidate
is required to be greater 
than~$100\mum/c$ 
and $150\mum/c$ 
for the~\Bc and \Bu~candidates, respectively,   
to~reduce the~background from 
particles originating directly from the~PV. 
To~suppress the~cross\nobreakdash-feed from
\decay{\Bp}{\jpsi\pip} decays, 
the~mass of the~$\jpsi\pip$ system for 
the~reconstructed \mbox{\BcTochicjpi} candidates 
is required to be outside of the~mass range
\mbox{$5.22<m_{\jpsi\pip}<5.35\gevcc$}. 

To further suppress 
%% large 
combinatorial background
for the~\BcTochicjpi, 
\BcTojpsipi 
and \mbox{$\decay{\Bu}{\chic\Kp}$}~decays,
three separate {\sc{BDTG}} classifiers are used, 
each trained on the~corresponding simulated samples.
As~a~proxy for the~background, 
the~\Bc~candidates from data with 
a~mass between 6.4 and 6.6\gevcc 
are used 
%% for \mbox{$\decay{\Bc}{\chic\pip}$} and 
%% \mbox{$\decay{\Bc}{\jpsi\pip}$}~classifiers,
for the~\Bc~classifiers
and \Bu~candidates
with a~mass between 5.35 and 5.50\gevcc
are used for 
%% the~\mbox{$\decay{\Bu}{\chic\Kp}$}~classifier.
the~\Bu~classifier. 
The~$k$\nobreakdash-fold
cross\nobreakdash-validation 
technique~\cite{geisser1993predictive} with $k = 13$ 
is used to avoid 
introducing a~bias in the~{\sc{BDTG}}~evaluation.  
The~{\sc{BDTG}} classifiers are trained
using variables related to the~reconstruction quality, 
decay time of \B~candidates,
kinematics of particles in the final state 
and variables related to hadron (pion for \Bc~candidates
and kaon for \Bu~candidates) 
identification~\cite{LHCb-PROC-2011-008,
LHCb-DP-2012-003}.
The~requirement on the~{\sc{BDTG}}~output is 
 chosen to maximise 
 the~Punzi figure\nobreakdash-of\nobreakdash-merit
 \mbox{$\upvarepsilon/(\upalpha/2+
 \sqrt{B})$}~\cite{Punzi:2003bu} 
 for \BcTochicjpi~decays 
 and $S/\sqrt{S+B}$  
 for 
 \BcTojpsipi 
 and \mbox{$\decay{\Bu}{\chic\Kp}$}~decays, 
 where 
 $\upvarepsilon$ is the~signal efficiency 
in simulation for \BcTochicjpi decays,
$\upalpha = 5$ 
 is the~desired signal significance
 in~units of standard deviations,  
 $B$~is the~expected background 
 yield  within a~narrow mass window
 centred at the~known 
 mass of the~$\bquark$\nobreakdash-hadron~\cite{PDG2023}, and 
 the~signal yields $S$ are estimated from simulated samples,
normalised to the~signal yields observed
in data for a loose requirement on the~{\sc{BDTG}}~output.
After application of the~{\sc{BDTG}}~requirement
a~small fraction of events, 
about 2.5\% for 
the~\mbox{$\decay{\Bc}{\chic\pip}$}
and \mbox{$\decay{\Bu}{\chic\Kp}$}~candidates, 
and 0.5\% for the~\mbox{$\decay{\Bc}{\jpsi\pip}$} candidates in 
the~signal region,  
contain multiple candidates.
For~each decay channel 
if two or more candidates 
are found 
in the~same event,
only one randomly chosen 
candidate is retained 
for further analysis.

\begin{figure}[tb]
\centering
{ \setlength{\unitlength}{1mm}
	\begin{picture}(150,122)
	%% 
        %%\graphpaper[5](-10,-10)(170,130)
        %%
    	\definecolor{root8}{rgb}{0.35, 0.83, 0.33}
		\put(0,2){\includegraphics*[width=150mm]{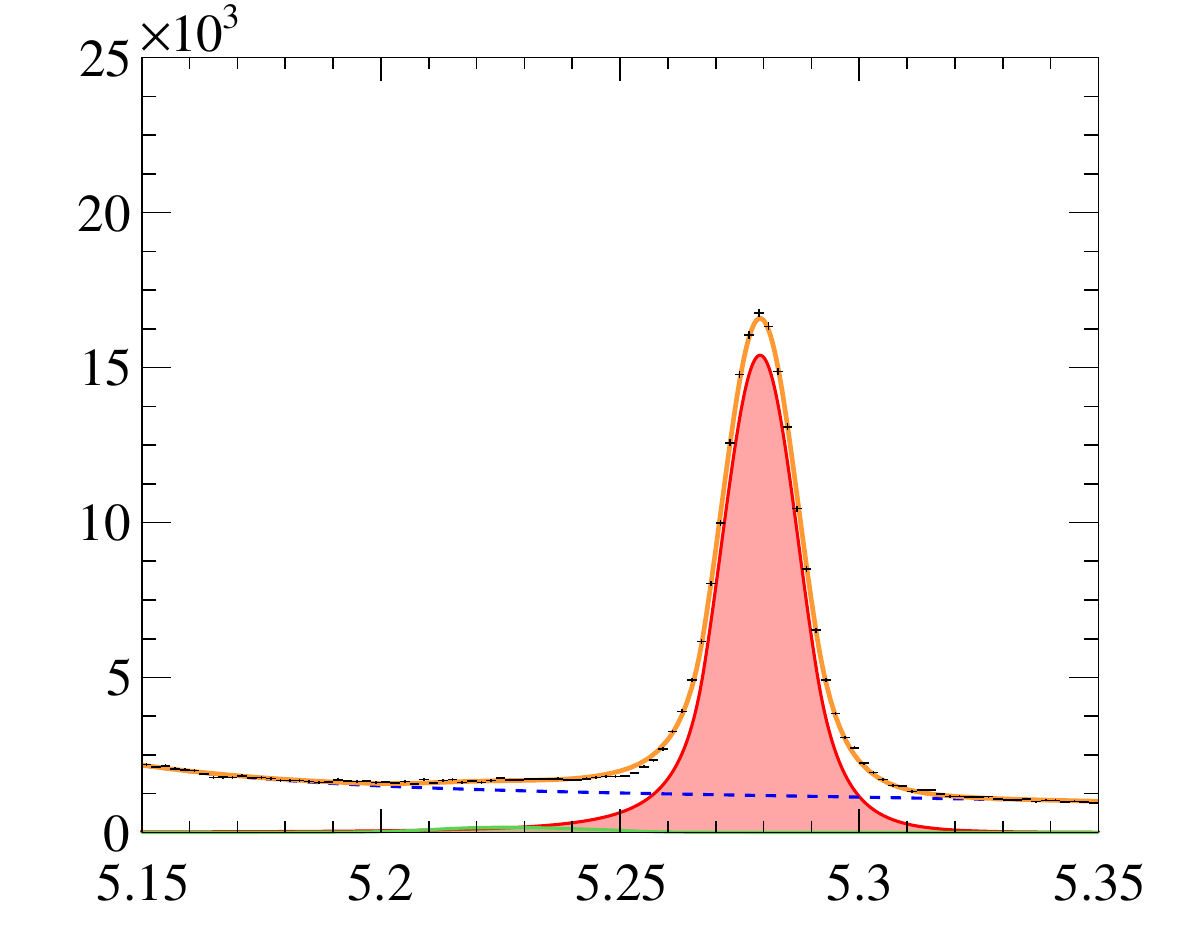}}
		\put(  0,48){\LARGE\begin{sideways}Candidates/$(2\mevcc)$\end{sideways}}
		\put( 74, 0){\LARGE$m_{\chic\Kp}$}
		\put(116, 0){\LARGE$\left[\!\gevcc\right]$}
		\put(112,100){\large$\begin{array}{l}\lhcb \\ 9\invfb\end{array}$}
	    \put(25,95){$\begin{array}{cl}
	    %% DATA 
	    \!\bigplus\mkern-5mu&\mathrm{data} 
	    \\ 
	    \begin{tikzpicture}[x=1mm,y=1mm]\filldraw[fill=red!35!white,draw=red,thick]  (0,0) rectangle (9,4);\end{tikzpicture} & \decay{\Bu}{\chicone\Kp} \\
	    \begin{tikzpicture}[x=1mm,y=1mm]\filldraw[fill=root8!35!white,draw=root8,thick]  (0,0) rectangle (9,4);\end{tikzpicture} & \decay{\Bu}{\chictwo\Kp} 
	    \\
	    {\color[RGB]{0,0,255}{\hdashrule[0.0ex][x]{8mm}{2.0pt}{1.0mm 0.4mm}}} & \mathrm{background}
	    \\
	    {\color[RGB]{255,153,51} {\rule{8mm}{2.0pt}}} & \mathrm{total}
	    \end{array}$}
	   \end{picture}
	}
	\caption{\small
	Mass distribution 
	for selected 
	\mbox{\decay{\Bu}{\chic\Kp}}~candidates
        with a~$\chicone$~mass\protect\nobreakdash-constraint. 
        The~result of the~fit, described in the~text, is overlaid.
	}
	\label{fig:signal_fit_bu}
\end{figure}

\begin{figure}[tb]
\centering
{ \setlength{\unitlength}{1mm}
	\begin{picture}(150,122)
	%% 
        %% \graphpaper[5](-10,-10)(170,130)
        %%
        \definecolor{root8}{rgb}{0.35, 0.83, 0.33}
        \definecolor{mag}{rgb}{255, 0, 255}
		\put(0,2){\includegraphics*[width=150mm]{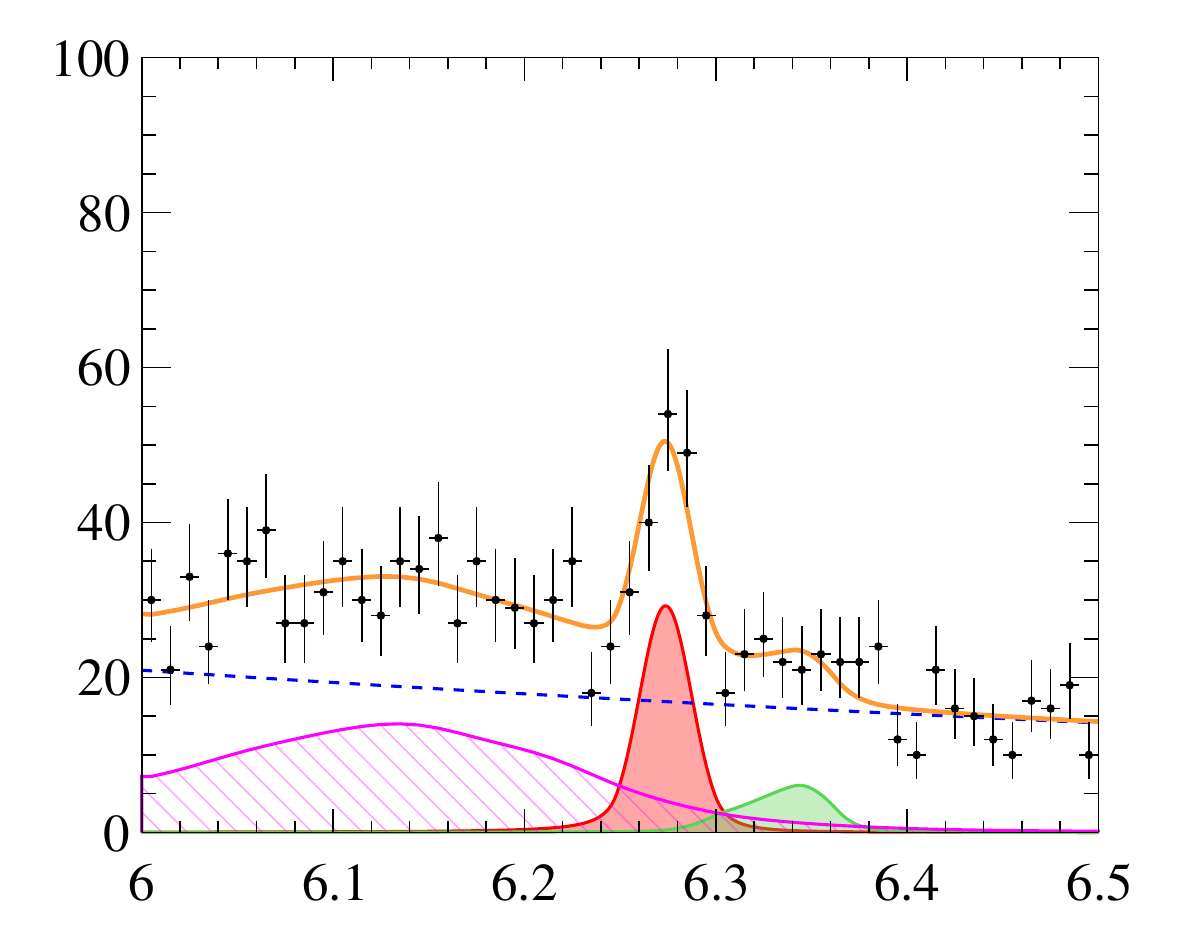}}
		\put( -2,46){\LARGE\begin{sideways}Candidates/$(10\mevcc)$\end{sideways}}
		\put( 74, 0){\LARGE$m_{\chic\pip}$}
		\put(116, 0){\LARGE$\left[\!\gevcc\right]$}
		\put(112,100){\large$\begin{array}{l}\lhcb \\ 9\invfb\end{array}$}
	    \put(25,90){$\begin{array}{cl}
	    %% DATA 
	    \!\bigplus\mkern-5mu&\mathrm{data} 
	    \\ 
	    \begin{tikzpicture}[x=1mm,y=1mm]\filldraw[fill=red!35!white,draw=red,thick]  (0,0) rectangle (9,4);\end{tikzpicture} & \decay{\Bc}{\chictwo\pip} \\
	    \begin{tikzpicture}[x=1mm,y=1mm]\filldraw[fill=root8!35!white,draw=root8,thick]  (0,0) rectangle (9,4);\end{tikzpicture} & \decay{\Bc}{\chicone\pip} 
	    \\
        \begin{tikzpicture}[x=1mm,y=1mm]\draw[ pattern={mylines[size=4pt,line width=.8pt,angle=-45]}, draw=mag, thick, pattern color=mag]  (0,0) rectangle (9,4);\end{tikzpicture} & \decay{\Bc}{\jpsi\pip\piz} 
        \\
	    {\color[RGB]{0,0,255}{\hdashrule[0.0ex][x]{8mm}{2.0pt}{1.0mm 0.4mm}}} & \mathrm{background}
	    \\
	    {\color[RGB]{255,153,51} {\rule{8mm}{2.0pt}}} & \mathrm{total}
	    \end{array}$}
	   \end{picture}
	}
	\caption{\small
	Mass distribution 
	for selected 
	\mbox{\BcTochicjpi}~candidates
   with a~$\chictwo$~mass constraint.
   The~result of the~fit, described in the~text, is overlaid.
	}
	\label{fig:signal_fit_chic}
\end{figure}

\begin{figure}[tb]
\centering
{ \setlength{\unitlength}{1mm}
	\begin{picture}(150,120)
	%% 
        %% \graphpaper[5](-10,-10)(170,130)
        %%
    	\definecolor{root8}{rgb}{0.35, 0.83, 0.33}
		\put( 0,2){\includegraphics*[width=150mm]{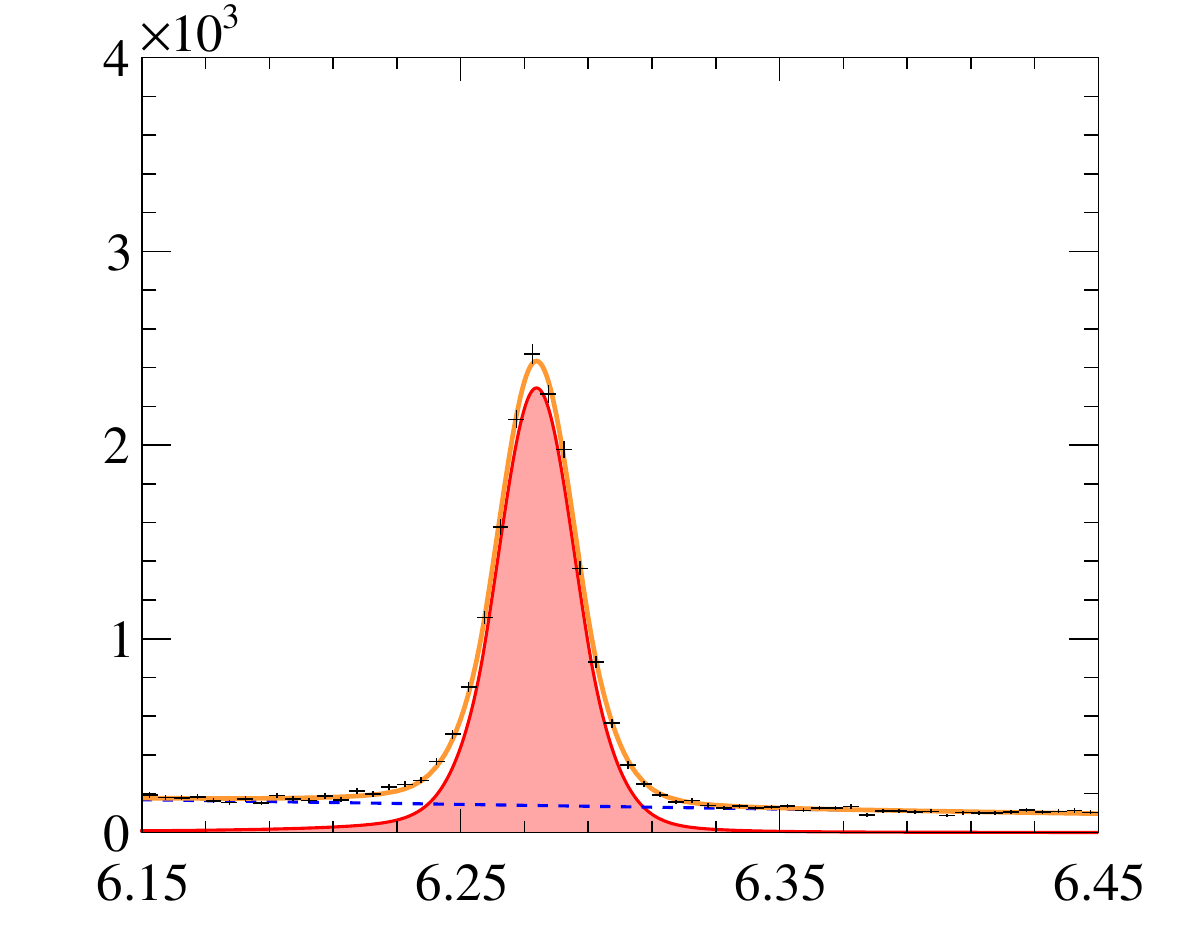}}
		\put(  2,48){\LARGE\begin{sideways}Candidates/$(5\mevcc)$\end{sideways}}
		\put( 74,0 ){\LARGE$m_{\jpsi\pip}$}
		\put(116,0 ){\LARGE$\left[\!\gevcc\right]$}
		\put(112,100){\large$\begin{array}{l}\lhcb \\ 9\invfb\end{array}$}
	 \put(25,95){$\begin{array}{cl}
	 %% DATA 
	 \!\bigplus\mkern-5mu&\mathrm{data} 
	 \\ 
	 \begin{tikzpicture}[x=1mm,y=1mm]\filldraw[fill=red!35!white,draw=red,thick]  (0,0) rectangle (9,4);\end{tikzpicture} & \decay{\Bc}{\jpsi\pip} 
	 \\
	    {\color[RGB]{0,0,255}{\hdashrule[0.0ex][x]{8mm}{2.0pt}{1.0mm 0.4mm}}} & \mathrm{background}
	 \\
	    {\color[RGB]{255,153,51} {\rule{8mm}{2.0pt}}} & \mathrm{total}
	 \end{array}$}

	   \end{picture}
	}
	\caption{\small
	Mass distributions 
	for selected 
	\mbox{\BcTojpsipi}~candidates.
    The~result of the~fit, 
    described in the~text, is overlaid.
	}
	\label{fig:signal_fit_jpsi}
\end{figure}

To improve the~\Bc and \Bu~mass resolution, 
the mass of the~$\Bc$ and $\Bu$~candidates
is calculated using
a~kinematic fit~\cite{Hulsbergen:2005pu}, 
similar to the~one described above, 
but with an~additional constraint
fixing the~mass of the~\jpsi\g
combination to the~known \chictwo~meson mass~\cite{PDG2023}
for the~\BcTochicjpi~candidates,
and to the~known mass of the~\chicone~meson~\cite{PDG2023}
for the~\mbox{$\decay{\Bu}{\chic\Kp}$}~candidates.
The~mass distributions for selected 
\mbox{$\decay{\Bu}{\chic\Kp}$},
\BcTochicjpi and \mbox{\BcTojpsipi} candidates 
are shown in Figs.~\ref{fig:signal_fit_bu}, 
\ref{fig:signal_fit_chic}~and~\ref{fig:signal_fit_jpsi}, respectively.

\section{Signal yields}
\label{sec:Sig_eff}

The yields  for the~\decay{\Bu}{\chic\Kp}, 
\mbox{\BcTochicjpi} and \BcTojpsipi~decays 
are determined 
using extended unbinned 
maximum\nobreakdash-likelihood fits.
In this analysis, the~detector resolution and a~possible 
mass bias for the~signal 
\mbox{$\decay{\Bc}{\chic\pip}$}~channel 
are determined from the~control channel 
\mbox{$\decay{\Bu}{\chic\Kp}$},
exploiting the~similarity between the~two final states.

The~fit model for the~\mbox{$\decay{\Bu}{\chic\Kp}$}~control channel 
consists of three components. 
The~first component corresponds to 
    the~decay \mbox{$\decay{\Bu}{\chicone\Kp}$},  
    located close to the~known mass
    of the~\Bu~meson,
    and is parameterised by a~modified Gaussian function 
	with power\nobreakdash-law tails on both sides of 
        the~distribution~\cite{LHCb-PAPER-2011-013,
	Skwarnicki:1986xj}.
The~second component corresponds to a~small
        contribution from \mbox{$\decay{\Bu}{\chictwo\Kp}$}~decays,
        and is parameterised by the~sum 
        of a~Gaussian function and 
        a~modified 
        Novosibirsk function~\cite{PhysRevD.84.112007}.
        Due to the~mass difference between 
        the~$\chicone$
        and $\chictwo$~states~\cite{E760:1991qim,
        Andreotti:2005ts,
        LHCb-PAPER-2017-036,
        PDG2023}, 
        the~$\chicone$\nobreakdash-mass constraint 
        used to evaluate the~mass of the~\Bu~candidates
        shifts  the~position of this component 
        toward  
        lower masses.  
The~third component corresponds to combinatorial background 
    and is modelled by the~product of an~exponential function
    and a~positive second\nobreakdash-order polynomial function~\cite{karlin1953geometry}.
The~shape parameters for the~fit components describing  
the~\mbox{$\decay{\Bu}{\chicone\Kp}$}
and \mbox{$\decay{\Bu}{\chictwo\Kp}$}~decay contributions
are taken from simulation and their uncertainties 
are propagated to the~fit using a~multivariate Gaussian constraint. 
The~position and resolution parameters for 
the~\mbox{$\decay{\Bu}{\chicone\Kp}$}~component 
are parameterised as 
\begin{subequations} \label{eq:params_Bu}
\begin{eqnarray}
  \upmu_{\Bu}     & = &  \updelta m_{\Bu} + m_{\Bu} \,,  
   \\ 
   \upsigma_{\Bu} & = &  s_{\Bu} \times  \upsigma_{\Bu}^{\mathrm{MC}} \,, 
\end{eqnarray} \label{eq:Bu_params}\end{subequations}
where $m_{\Bu}$~is the~known mass of the~\Bu~meson and  
$\upsigma_{\Bu}^{\mathrm{MC}}$~is a~resolution parameter
obtained from simulation. 
The~parameter 
$\updelta m_{\Bu}$ 
accounts for a~mass bias,
while
$s_{\Bu}$
describes potential
differences between data and 
simulation for the~resolution.
The~position parameters
for 
the~\mbox{$\decay{\Bu}{\chicone\Kp}$}
and \mbox{$\decay{\Bu}{\chictwo\Kp}$}~components
are allowed to vary in the~fit,
but 
their~difference 
is constrained 
to the~value obtained from simulation
using a~Gaussian constraint.
The~resulting fit is overlaid
in Fig.~\ref{fig:signal_fit_bu}, and 
the~parameters of interest are listed 
in Table~\ref{tab:fit_res_Bu}. 
The~yield of 
the~\mbox{$\decay{\Bu}{\chictwo\Kp}$}~decay
is found to be much smaller
than that of 
the~\mbox{$\decay{\Bu}{\chicone\Kp}$}~decay
as seen in previous measurements~\cite{Belle:2008qeq, 
Aubert:2008ae,
LHCb-PAPER-2013-024} 
and in accordance with expectations
from QCD~factorisation. 
The~obtained
mass bias parameter $\updelta m_{\Bu}$
and resolution scale parameter $s_{\Bu}$
are used for the~determination of 
the~signal yield
for the~\mbox{$\decay{\Bc}{\chic\pip}$}~channel.

\begin{table}[b]
	\centering
	\caption{\small 
        Parameters of interest 
	from the~fit
          to the~$\chic\Kp$~mass spectrum:
          the~yields of 
          the~\mbox{$\decay{\Bu}{\chicone\Kp}$}
          and \mbox{$\decay{\Bu}{\chictwo\Kp}$}~decays, 
          the~\Bu~mass bias parameter
          $\updelta m_{\Bu}$ and 
          the~resolution scale factor 
            $s_{\Bu}$.
	}
       \label{tab:fit_res_Bu}
	\begin{tabular}{llc}
     \multicolumn{2}{l}{Parameter}  & Value  
    \\[1.5mm]
    \hline     
    \\[-3.5mm]
$N_{\decay{\Bu}{\chicone\Kp}}$ & $\left[10^3\right]$  & $\phantom{-}171.35 \pm 0.61\phantom{00} $  
\\
$N_{\decay{\Bu}{\chictwo\Kp}}$ & $\left[10^3\right]$& $ \phantom{-}3.28 \pm 0.40 $  
\\
$\updelta m_{\Bu}$ & $\left[\!\mevcc\right]$ 
%% & $-0.004\pm0.124$
& $\phantom{-}0.07\pm0.12$
\\
$s_{\Bu}$  &   &   $\phantom{-}1.102\pm0.004$ 
\end{tabular}
\end{table}

The~fit model for the~\BcTochicjpi~channel consists of four components. 
The~first component corresponds to the~\BcTochictwopi~decay and 
is parameterised by a~modified Gaussian
function.
This component is located close to 
the~known mass of the~\Bc~meson.
The~second smaller component corresponds 
to the~\BcTochiconepi~decay
and is parameterised by the~sum of~a~modified Gaussian
and a~modified Novosibirsk function.
Due~to the~mass difference between
the~$\chicone$ and $\chictwo$~states,   
and the~$\chictwo$\nobreakdash-mass constraint 
used to evaluate the~mass of the~\Bc~candidate,
the~position of this component
is shifted towards higher masses.
The~third component corresponds to the~partially 
reconstructed \mbox{$\decay{\Bc}{\jpsi\pip 
\left(\decay{\piz}
{\g\g}\right)}$}~decay~\cite{LHCb-PAPER-2023-046}
where 
one photon 
%% is not reconstructed and another
is used in the~misreconstructed 
\mbox{$\decay{\chic}{\jpsi\g}$}~candidate
and the~other is not reconstructed. 
The~shape of this component is obtained from 
the~simulation. %%  of \mbox{$\decay{\Bc}{\jpsi\pip\piz}$} decays.
Finally, the~combinatorial background is parameterised 
by an~exponential function.
Similar to the~\mbox{$\decay{\Bu}{\chic\Kp}$}~case above, 
the~shape parameters for the~fit components describing  
the~\mbox{$\decay{\Bc}{\chicone\pip}$}
and \mbox{$\decay{\Bc}{\chictwo\pip}$}~contributions
and the~difference between the~positions of 
the~mass peaks 
of 
the~\mbox{$\decay{\Bc}{\chicone\pip}$}
and \mbox{$\decay{\Bc}{\chictwo\pip}$}~contributions
are taken from simulation. 
Their uncertainties 
are propagated to the~fit using 
a~multivariate Gaussian constraint. 
Similar to Eq.~\refeq{eq:params_Bu}, 
the~position and resolution parameters for 
the~\mbox{$\decay{\Bc}{\chictwo\pip}$}~component 
are parameterised~as 
\begin{subequations}\label{eq:params_Bc}
\begin{eqnarray}
   \upmu_{\Bc}     & = &  \updelta m_{\Bu} + m_{\Bc} \,,  
   \\ 
   \upsigma_{\Bc} & = &  s_{\Bu} \times  
   \upsigma_{\Bc}^{\mathrm{MC}} \,, 
\end{eqnarray}
\end{subequations}
where $m_{\Bc}$~is the~known 
mass of the~\Bc~meson~\cite{LHCb-PAPER-2020-003} and 
$\upsigma_{\Bc}^{\mathrm{MC}}$~is a~resolution parameter
obtained from simulation. 
The~values of $\updelta m_{\Bu}$ 
and $s_{\Bu}$ 
are defined in Eq.~\eqref{eq:Bu_params}
and are taken
from the~fit to the~$\chicone\Kp$~mass spectrum above
(see Table~\ref{tab:fit_res_Bu}).
Their uncertainties, 
as well as the~uncertainty for 
the~\mbox{$m_{\Bc}$}~value, 
are propagated to 
the~fit using Gaussian constraints. 
%%
%% {\color{red}{
The~yield of 
the~\BcTochiconepi~signal component,  
$N_{\decay{\Bc}{\chicone\pip}}$, 
is parameterised~as
\begin{subequations}
\begin{equation}
\label{eq:yield_chic1}
    N_{\decay{\Bc}{\chicone\pip}} \equiv 
    N_{\decay{\Bc}{\chictwo\pip}} 
    \times 
    \mathcal{R}^{\chicone}_{\chictwo}
    \times
    \upvarepsilon^{\chicone}_{\chictwo}    %% \dfrac
    %% { \upvarepsilon_{\decay{\Bc}{\chicone\pip}}}
    %% { \upvarepsilon_{\decay{\Bc}{\chictwo\pip}}}   
    \times    
          \dfrac{ \BR_{\decay{\chicone}{\jpsi\g}}}
                { \BR_{\decay{\chictwo}{\jpsi\g} }}  \,, 
\end{equation}
where
$N_{\decay{\Bc}{\chictwo\pip}}$
is
the~yield
of the~\BcTochictwopi~signal 
component,   
$\mathcal{R}^{\chicone}_{\chictwo}$ is 
a~free parameter in the~fit corresponding to 
the~ratio of branching fractions
for the~$\decay{\Bc}{\chicone\pip}$
and $\decay{\Bc}{\chictwo\pip}$~decays 
% \begin{subequations}
\begin{equation}
\mathcal{R}^{\chicone}_{\chictwo} 
\equiv 
\dfrac{\BR_{\BcTochiconepi}}{\BR_{\BcTochictwopi}}\,,  \label{eq:f12}
\end{equation}
$\BR_{\decay{\chic}{\jpsi\g}}$ 
are the~branching 
fractions for the~\mbox{$\decay{\chic}{\jpsi\g}$}~decays~\cite{PDG2023}, 
and
$\upvarepsilon^{\chicone}_{\chictwo}$
is the~ratio of the~total efficiencies
for the~\mbox{\BcTochictwopi} and \mbox{\BcTochiconepi}~decays (see Sec.~\ref{sec:ratios})
\begin{equation}
    \upvarepsilon^{\chicone}_{\chictwo} \equiv 
    \dfrac
    { \upvarepsilon_{\decay{\Bc}{\chicone\pip}}}
    { \upvarepsilon_{\decay{\Bc}{\chictwo\pip}}}\,.  \label{eq:e12}
\end{equation}
\end{subequations}
The~uncertainties for
%% the~$\mathcal{R}^{\chicone}_{\chictwo}$ 
the~\mbox{$\BR_{\decay{\chic}{\jpsi\g}}$}
and $\upvarepsilon^{\chicone}_{\chictwo}$~values 
%% these values, 
including the~systematic effects
on the~ratio of efficiencies, 
discussed in detail in Sec.~\ref{sec:Systematics}, 
%% related to the efficiency measurement in Eq.~\ref{eq:yield_chic1}
are 
%% propagated 
accounted for in the~fit via Gaussian constraints. 
The~result of the~fit is overlaid
in Fig.~\ref{fig:signal_fit_chic}.
The~parameters of interest, namely
the~yield of the~\mbox{$\decay{\Bc}{\chictwo\pip}$}~decays
and 
the~ratio of branching fractions
$\mathcal{R}^{\chicone}_{\chictwo}$, 
are found to be 
%% \begin{subequations}
\begin{eqnarray*}
N_{\decay{\Bc}{\chictwo\pip}} & = & 108 \pm 16 \,, \label{eq:N_chictwo}
\\
\mathcal{R}^{\chicone}_{\chictwo} & = & 0.24^{\,+\,0.13}_{\,-\,0.11} \,.
\end{eqnarray*}    
%% \end{subequations}
The~significance 
for the~\BcTochictwopi~signal 
is estimated using Wilks' theorem~\cite{Wilks:1938dza}
and found to be 8.1~standard deviations,
corresponding to the~first observation 
of the~\mbox{\BcTochictwopi}~decay.
On~the~contrary, 
the~\BcTochiconepi~signal 
is found 
to be insignificant,
exhibiting 
a~hierarchy 
of branching fractions 
opposite to that in the~\mbox{$\decay{\Bu}{\chic\Kp}$}~case,
but in qualitative 
agreement with 
the~theory expectation 
for the~suppression of 
%the~decays via 
the~$\chicone$~state 
in ~\mbox{$\decay{\Bc}{\chic\pip}$}~decays~\cite{
PhysRevD.65.014017,
PhysRevD.82.034019,
PhysRevD.74.074008,
PhysRevD.73.054024,
vvKiselev_2002,
PhysRevD.97.033001,
Wang_2012,
ZHU2018359,
Chao-Hsi_2001}. 
Using the~$\mathrm{CL_s}$ technique~\cite{CLs},  
where the~$p$\nobreakdash-values are 
calculated based on the~asymptotic 
properties of the~profile likelihood 
ratio~\cite{Cowan:2010js}, 
an~upper limit of
\begin{equation}
\mathcal{R}^{\chicone}_{\chictwo} < 0.41   \label{eq:f12_stat}
\end{equation}
is found at 90\% confidence level.

The~fit model for the~normalisation \BcTojpsipi channel consists of two components.  
The~first component corresponds to the~\BcTojpsipi~decay
   and is  parameterised by the~sum 
   of  
   a~modified Gaussian function
   and a~Gaussian
   function with a~common mean value.
The~second component describes the~combinatorial background 
    and is parameterised by 
    a~positive first\nobreakdash-order polynomial function~\cite{karlin1953geometry}.
Similar to 
the~\mbox{$\decay{\Bu}{\chic\Kp}$}
and \mbox{$\decay{\Bc}{\chic\pip}$}~cases above, 
the~shape parameters for the~fit component describing  
the~\mbox{$\decay{\Bc}{\jpsi\pip}$}~contributions
are taken from simulation and their uncertainties 
are propagated to the~fit using a~multivariate 
Gaussian constraint. 
The~resolution parameter for 
the~\mbox{$\decay{\Bc}{\jpsi\pip}$}~component 
is parameterised as 
%% \begin{subequations}
\begin{equation}
   \upsigma_{\Bc}  = s_{\Bc} \times  
   \upsigma_{\Bc}^{\mathrm{MC}} \,, 
\end{equation}
%% \end{subequations}
where
$\upsigma^{\Bc}_{\mathrm{MC}}$~is 
a~resolution parameter
obtained from simulation
and the~parameter 
%% of interest 
$s_{\Bc}$
accounts for a~possible 
difference between data and 
simulation for the~resolution parameter.
The~resulting fit is overlaid
in Fig.~\ref{fig:signal_fit_jpsi}.
The~parameters of interest, namely 
the~yield of the~\mbox{$\decay{\Bc}{\jpsi\pip}$}~signal
and the~resolution scale parameter $s_{\Bc}$, 
are found to be 
%% \begin{subequations}
\begin{eqnarray*}
N_{\decay{\Bc}{\jpsi\pip}} & = & \left( 15.40 \pm 0.14 \right)\times 10^3 \,,
\label{eq:N_jpsi}
\\
s_{\Bc}  & = & \ 1.106 \pm 0.011 \,.
\end{eqnarray*}
%%\end{subequations}

%%
To~validate the~observed 
\mbox{$\decay{\Bc}{\chictwo\pip}$}~signal, 
several cross\nobreakdash-checks are performed. 
The~data are categorised according 
to the~data\nobreakdash-taking 
period,
the~polarity of the~LHCb dipole  magnet~\cite{LHCb-TDR-001}
and the~charge of the~\Bc~candidates. 
Also, the~\chictwo~mass constraint, used for calculation 
of the~mass of the~\Bc~candidates,
is replaced by a~\chicone~mass constraint.
The~results are found to be consistent 
among all samples and analysis techniques.

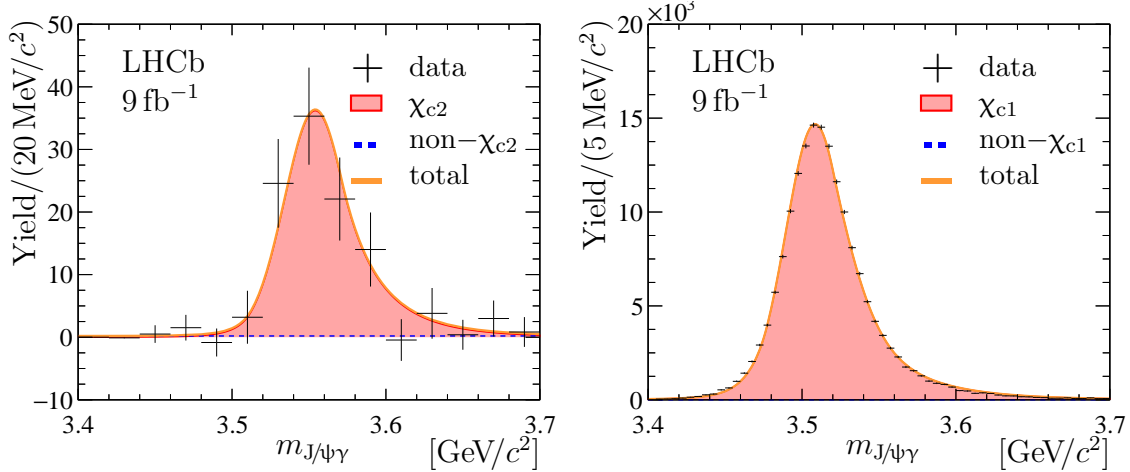
\begin{figure}[tb]
\centering
{ \setlength{\unitlength}{1mm}
\begin{picture}(150,60)
	%% 
        %% \graphpaper[5](-10,-10)(170,130)
        %%
        \definecolor{root8}{rgb}{0.35, 0.83, 0.33}
		\put( 0, 0){\includegraphics*[width=75mm]{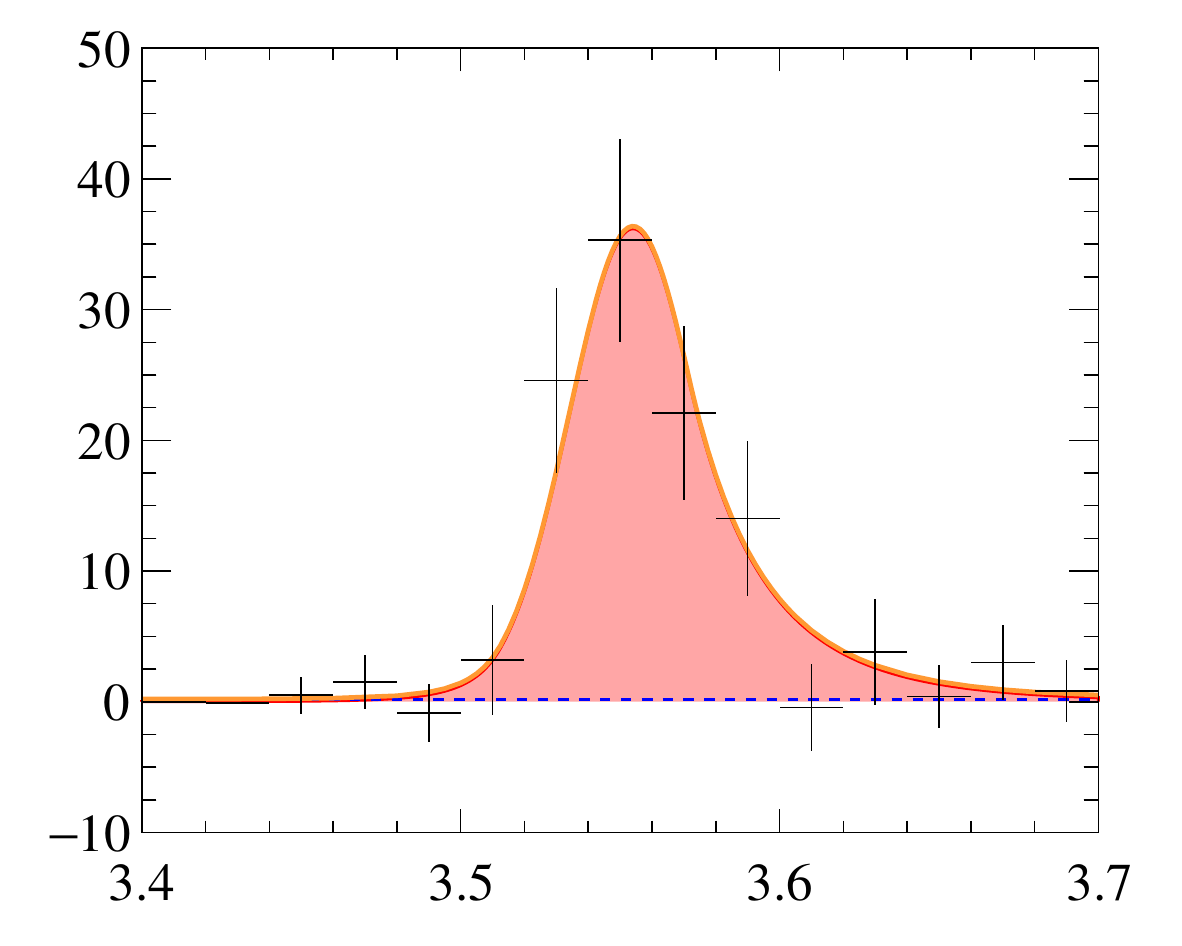}}
  		\put(75, 0){\includegraphics*[width=75mm]{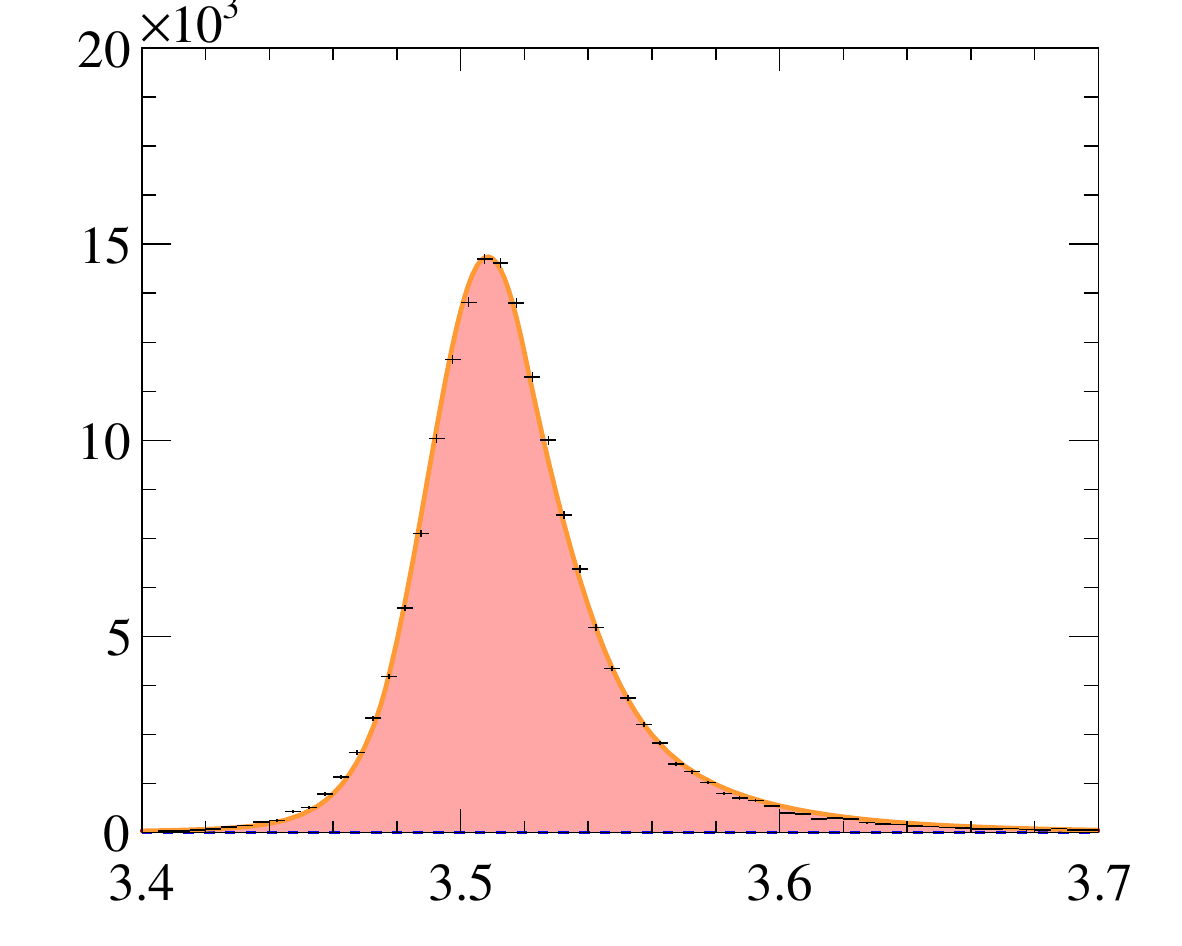}}
		\put(0,25){\begin{sideways}Yield/$(20\mevcc)$\end{sideways}}
            \put(75,27){\begin{sideways}Yield/$(5\mevcc)$\end{sideways}}
  
		\put( 35,0){$m_{\jpsi\g}$}
            \put(110,0){$m_{\jpsi\g}$}
            \put( 55,-1) {$\left[\!\gevcc\right]$}
            \put(130,-1){$\left[\!\gevcc\right]$}
   
    \put(43,43){\small$\begin{array}{cl}
	 %% DATA 
	 \!\bigplus\mkern-5mu&\mathrm{data} 
	 \\ 
	 \begin{tikzpicture}[x=1mm,y=1mm]\filldraw[fill=red!35!white,draw=red,thick]  (0,0) rectangle (4,2);\end{tikzpicture} 
      & \chictwo 
	 \\
     {\color[RGB]{0,0,255}{\hdashrule[0.0ex][x]{4mm}{2.0pt}{1.0mm 0.4mm}}} 
     & 
     %% \cancel{\chictwo}
     \mathrm{non-}\chictwo
	 \\
      {\color[RGB]{255,153,51} {\rule{4mm}{2.0pt}}} & \mathrm{total}
	 \end{array}$}
     \put(118,43){\small$\begin{array}{cl}
	 %% DATA 
	 \!\bigplus\mkern-5mu&\mathrm{data} 
	 \\ 
	 \begin{tikzpicture}[x=1mm,y=1mm]\filldraw[fill=red!35!white,draw=red,thick]  (0,0) rectangle (4,2);\end{tikzpicture} 
      & \chicone 
	 \\
     {\color[RGB]{0,0,255}{\hdashrule[0.0ex][x]{4mm}{2.0pt}{1.0mm 0.4mm}}} 
     & 
     %% \cancel{\chicone}
     \mathrm{non-}\chicone
	 \\
      {\color[RGB]{255,153,51} {\rule{4mm}{2.0pt}}} & \mathrm{total}
	 \end{array}$}
    \put(13,48){$\begin{array}{l}\lhcb \\ 9\invfb\end{array}$}
    \put(88,48){$\begin{array}{l}\lhcb \\ 9\invfb\end{array}$}
\end{picture}
}
\caption{\small
       Background-subtracted 
       $\jpsi\g$~mass distribution
       from
       (left)~the~\mbox{\BcTochictwopi}~decays
       and 
       (right)~\mbox{$\decay{\Bu}{\chicone\Kp}$}~decays.
       Projections of the~fit, described in the~text, are overlaid. 
	}
	\label{fig:signal_chictwo}
\end{figure}

To confirm the~\mbox{$\decay{\chictwo}{\jpsi\g}$}~signal 
within the~observed \mbox{$\decay{\Bc}{\chictwo\pip}$} signal, 
the~background\nobreakdash-subtracted $\jpsi\g$~mass spectrum is examined. 
For~background subtraction, the~\sPlot technique \cite{Pivk:2004ty} 
%% based on the~fit result described earlier, 
is used
with the~$\chic\pip$~mass 
as the~discriminating variable. 
The~background\nobreakdash-subtracted $\jpsi\g$~spectrum is shown in 
Fig.~\ref{fig:signal_chictwo}\,(left).
The~\mbox{$\decay{\chicone}{\jpsi\g}$}~signal 
from the~control sample of   \mbox{$\decay{\Bu}{\chicone\Kp}$}~decays
is used to examine the~possible bias in the~mass 
and mass resolution for the~$\jpsi\g$~mass. 
The~background\nobreakdash-subtracted 
$\jpsi\g$~spectrum 
from the~\mbox{$\decay{\Bu}{\chicone\Kp}$}~decays
is presented in 
Fig.~\ref{fig:signal_chictwo}\,(right). 
Unbinned extended fits are performed
to the~$\jpsi\g$ distributions 
with a~model consisting  of two components.
The~first component corresponds to 
          the~\mbox{$\decay{\chic}{\jpsi\g}$}~signal 
          and is
          parameterised with  a~modified Gaussian
          function for the~\chictwo~state and the~sum of a~modified Gaussian function
          and a~Gaussian function with a~common mean value for the~\chicone~state.
The~second component describes 
         the~\mbox{$\decay{\Bc}{\jpsi\g\pip}$}
         and \mbox{$\decay{\Bu}{\jpsi\g\Kp}$}~decays
         without intermediate $\chictwo$ and $\chicone$~mesons
         and is parameterised with a~constant.
Similar to the~fits described above, 
the~shape parameters for the~$\chic$~components
are taken from simulation and their uncertainties 
are propagated to the~fits using a~multivariate Gaussian constraint. 
Similar to Eqs.~\eqref{eq:params_Bu} and~\eqref{eq:params_Bc}
the~position and resolution parameters for 
each $\chic$~component in the~fits 
are parameterised~as 
\begin{subequations}
\begin{eqnarray}
   \upmu_{\chic}    & = &  \updelta m_{\chicone} + m_{\chic} \,,  
   \\ 
   \upsigma_{\chic} & = &  s_{\chicone} \times  
   \upsigma_{\chic}^{\mathrm{MC}} \,, 
\end{eqnarray}
\end{subequations}
where $m_{\chic}$~stands for the~known mass of the~corresponding 
\chic~state
and 
$\upsigma^{\chic}_{\mathrm{MC}}$~denotes the~corresponding 
resolution parameter obtained from simulation. 
The~mass~bias parameter
\mbox{$\updelta m_{\chicone}$}
and resolution scale parameter
\mbox{$s_{\chicone}$},
determined from the~fit 
to~the~$\jpsi\g$~mass spectrum 
from the~control \mbox{$\decay{\Bu}{\chicone\Kp}$}~channel, 
are
\mbox{$\updelta m_{\chicone}=-2.09\pm0.09\mevcc$} 
and \mbox{$s_{\chicone}=1.027\pm0.004$}. 
These values are used in
the~fit to the~$\jpsi\g$~mass spectrum 
for 
the~signal 
\mbox{$\decay{\Bc}{\chictwo\pip}$}~channel 
and their~uncertainties are included 
using Gaussian constraints. 
The~results of the~fits
to both~$\jpsi\g$~mass spectra are overlaid 
in Fig.~\ref{fig:signal_chictwo}.
For~both \mbox{$\decay{\Bu}{\chicone\Kp}$}
and \mbox{$\decay{\Bc}{\chictwo\pip}$}~decays 
a~good description of data is achieved 
without contributions from 
the~non-$\chic$~components.
This~proves 
the~observed 
\mbox{$\decay{\Bc}{\left(\jpsi\g\right)\pip}$}~signal 
proceeds via the~intermediate \mbox{$\decay{\chictwo}{\jpsi\gamma}$}~decay.

\section{Efficiency and ratio of branching fractions}
 \label{sec:ratios}

For each channel, the~efficiency is defined 
as the~product of the~detector acceptance, reconstruction,
selection and trigger efficiencies, 
where each subsequent efficiency 
is defined with
respect to the~previous one. 
Each~of the~efficiencies is calculated using 
the~simulation samples
described in Sec.~\ref{sec:Detector}.
The~ratio of efficiencies 
for the~\mbox{$\decay{\Bc}{\chicone\pip}$} and 
\mbox{$\decay{\Bc}{\chictwo\pip}$}~channels,
$\upvarepsilon^{\chicone}_{\chictwo}$,
defined in Eq.~\eqref{eq:e12}, 
and the~ratio of efficiencies
for the~\mbox{$\decay{\Bc}{\chictwo\pip}$} and 
\mbox{$\decay{\Bc}{\jpsi\pip}$}~channels,
$\upvarepsilon^{\chictwo}_{\jpsi}$, 
are found to be 
\begin{subequations}
\label{eq:effs}
    \begin{eqnarray}
 \upvarepsilon^{\chicone}_{\chictwo} & = & 
 \ 0.763 \pm 0.003 \,,
 \\
 \upvarepsilon^{\chictwo}_{\jpsi} & = &  
 \left( 9.84\phantom{,}\, \pm 0.03\phantom{0} \right) \times 10^{-2}\,, \label{eq:e2j}
    \end{eqnarray}
\end{subequations}
where the~uncertainties are due to the~size of simulated samples.

The~ratio of branching fractions
for the~\BcTochictwopi and 
\BcTojpsipi~decays is
calculated as 
\begin{equation}
     \mathcal{R}^{\chictwo}_{\jpsi}
     \equiv 
\dfrac{   \BR_{{\BcTochictwopi}}  } 
{   \BR_{{\BcTojpsipi}}  }  
     = 
     \dfrac{ N_{\mathrm{\BcTochictwopi}}}
           { N_{\mathrm{\BcTojpsipi}}} 
    \times 
     \dfrac {1} { \upvarepsilon^{\chictwo}_{\jpsi} }  
    \times\dfrac{1}{\BR_{\decay{\chictwo}{\jpsi\g}}}\,, \label{eq:br_rat}
\end{equation}
where the~signal yields $N$
are obtained in Sec.~\ref{sec:Sig_eff},  
$\upvarepsilon^{\chictwo}_{\jpsi}$
is the~efficiency ratio from Eq.~\eqref{eq:e2j} 
and   
$\BR_{\decay{\chictwo}{\jpsi\g}}$ is the~known branching
fraction of the~\decay{\chictwo}{\jpsi\g} decay~\cite{PDG2023}.
The~ratio of branching fractions is found to be 
\begin{equation*}
\mathcal{R}^{\chictwo}_{\jpsi} = 
%% 0.374\pm0.058  \,, 
0.37 \pm 0.06 \,, 
\end{equation*}
where the~uncertainty is statistical only. 
Systematic uncertainties are discussed 
in the~following section.

\section{Systematic uncertainties}
\label{sec:Systematics}

The~decay channels under study 
have similar kinematics
and topologies; therefore,
many sources of systematic uncertainty
cancel 
in~the~branching fraction
ratios. 
The~remaining contributions 
to the~systematic uncertainty are 
summarised in Table~\ref{tab:systematics} 
and discussed below.

An~important source of systematic uncertainty 
%% on the~ratios
is the~imperfect knowledge 
of the~shapes of the~signal and background 
components used in the~fits. 
To~estimate this, several alternative shapes 
are tested.  
For~the~\BcTochictwopi and \BcTojpsipi~signal shapes
two alternative models are tested:
\begin{itemize}
\item 
the~sum of a~generalised Student's
$t$\nobreakdash-distribution~\cite{Jackman} 
and a~Gaussian function; 
\item 
the~sum of  a~modified 
Apollonios function~\cite{Santos:2013gra} 
and a~Gaussian function.
\end{itemize}
For the~\BcTochiconepi~signal shape two alternative models are probed:
\begin{itemize}
    \item 
    the~sum of a~generalised Student's
    $t$\nobreakdash-distribution
    and a~modified Novosibirsk function~\cite{PhysRevD.84.112007}; %% ~\cite{PhysRevD.84.112007}; 
   \item 
    the~sum of a~modified Apollonios
    function
    and a~modified Novosibirsk function. 
\end{itemize}
For~the~background components, 
positive convex decreasing 
polynomial functions of the~second and 
third\nobreakdash-order 
are used as alternative background shapes
for the~fit to the~$\chic\pip$~mass spectrum.
For the~fit to the~$\jpsi\pip$~mass spectrum 
the~product of an~exponential function 
and a~positive first\nobreakdash-order 
polynomial function 
as well as a~positive second\nobreakdash-order polynomial function 
are tested. 
The~systematic uncertainty related to the~fit model is estimated using 
pseudoexperiments produced with the~baseline fit model and fit with 
sets of the~alternative models.
The~maximal deviations 
in the~ratios of 
the~signal yields
for 
the~\mbox{$\decay{\Bc}{\chictwo\pip}$}
and \mbox{$\decay{\Bc}{\jpsi\pip}$}~decays
with  respect to the~baseline models
do not exceed 0.3\% 
%% for the~variations 
for the~alternative signal models
and  0.7\% for 
the~alternative background models.
These values are taken as systematic uncertainties
for the~ratio 
%% $\mathcal{R}^{\BcTochictwopi}_{\BcTojpsipi}$.
$\mathcal{R}^{\chictwo}_{\jpsi}$.

\begin{table}[t]
	\centering
	\caption{\small
    Relative systematic uncertainties (in \%) for 
    the~ratio of branching fractions
    $\mathcal{R}^{\chictwo}_{\jpsi}$
    and 
    the~ratio of efficiencies 
    for the~\mbox{$\decay{\Bc}{\chicone\pip}$}
    and \mbox{$\decay{\Bc}{\chictwo\pip}$}~decays.
    The~total uncertainty is 
    the~quadratic sum of individual contributions.} 
	\label{tab:systematics}
	\vspace{2mm}
	\begin{tabular*}{0.70\textwidth}
	{@{\hspace{5mm}}l@{\extracolsep{\fill}}c@{\hspace{5mm}}c@{\hspace{5mm}}}
	Source  
        &  $\mathcal{R}^{\chictwo}_{\jpsi}$ 
        & 
        $\upvarepsilon^{\chicone}_{\chictwo}$
   \\[2.5mm]
  \hline 
  \\[-3.5mm]
  Fit model                  &   
  \\
  ~~~Signal shape            & $   0.3$  &   --- %% $   1.0$ 
  \\
  ~~~Background shape        & $   0.7$  &   --- %% $   3.7$ 
  \\
  Multiple candidates exclusion& $2.0$   &   --- %% $   2.2$ 
  \\
  \Bc~production spectra     & $  0.3$   & $   0.1$ 
  \\
  Track reconstruction     & $  <0.1\phantom{00}$&$  <0.1\phantom{00}$
  \\
  Photon reconstruction  & $  3.0$ &    ---
  \\
  Pion identification  & $0.5$  & $0.6$
  \\ 
  Trigger efficiency         & $  1.1$  & $  1.1$ 
  \\
  Data-simulation difference & $  2.1$  & ---
  \\
  Size of simulated sample   &  $ 0.3$   &  $ 0.4$ 
    \\[1.5mm]
  \hline 
  \\[-3.5mm]
  Total   &   $4.4$ &   1.3%% $4.6$ 
 
	\end{tabular*}
	\vspace{3mm}
\end{table}

In~the~analysis, 
multiple 
candidates are randomly excluded. 
To~estimate the~systematic effect due to 
the~exclusion procedure 
the~analysis is repeated 300~times
with different sets of candidates 
randomly removed.
The~variation of the~relative signal yield 
between \BcTochictwopi and \BcTojpsipi channels
%% in these pseudoexperiments
is found to be 2.0\%
and this value is assigned as 
the~corresponding systematic uncertainty.

An~important systematic uncertainty arises 
from differences 
%% in the~distributions 
%% in
between 
data and simulation. 
The~transverse momentum and rapidity spectra of 
the~\Bc mesons in the~simulated samples 
are adjusted to match those observed 
in a~high\nobreakdash-yield, 
low\nobreakdash-background
sample of \BcTojpsipi decays. 
The~finite size of this sample causes 
uncertainty in the~obtained production
spectra of the~\Bc~mesons. 
The~associated systematic uncertainty in the~efficiency
ratios
is estimated using the~variation of 
the~\Bc~kinematic spectra 
within their uncertainties.
The~induced variations 
for the~ratio 
$\mathcal{R}^{\chictwo}_{\jpsi}$
and for the~efficiency ratio 
$\upvarepsilon^{\chicone}_{\chictwo}$
are found to be 0.3\% and 0.1\%, respectively.
These values are taken as 
corresponding systematic uncertainties. 

Due to the~slightly different kinematic distributions of
the~final\nobreakdash-state particles
there are residual differences in 
the~reconstruction 
efficiency of charged\nobreakdash-particle tracks
that do not cancel completely in the~efficiency ratios.
The~track\nobreakdash-finding 
efficiency obtained from simulated samples
is corrected using calibration 
channels~\cite{LHCb-DP-2013-002}. 
The~uncertainties related to the~efficiency
correction factors are propagated to the~ratios 
of the~total 
efficiencies using pseudoexperiments
and found to be smaller than 0.1\%
both for 
 the~ratio 
$\mathcal{R}^{\chictwo}_{\jpsi}$
and for the~efficiency ratio 
$\upvarepsilon^{\chicone}_{\chictwo}$.
Differences in the~photon reconstruction efficiencies
between data and simulation are studied using 
a~large sample of 
\mbox{$\decay{\Bu}{\jpsi\left(\decay{\Kstarp}{\Kp\piz}\right)}$}
decays~\cite{
%% LHCb-PAPER-2012-022, 
%% LHCb-PAPER-2012-053, 
Govorkova:2015vqa, 
Govorkova:2124605, 
Belyaev:2016cri}.
The~uncertainty due to 
the~finite size of the~sample is propagated 
to the~ratio of the~total efficiencies using
pseudoexperiments and is found to be less than~0.1\%.
The additional uncertainty due to the~imprecise knowledge 
of the~ratio of branching fractions for the~\mbox{$\decay{\Bu}{\jpsi\Kstarp}$}
and~\mbox{$\decay{\Bu}{\jpsi\Kp}$}~decays~\cite{PDG2023}
is 3.0\%. The~squared sum of these uncertainties 
is taken as a~systematic 
uncertainty for the~ratio 
$\mathcal{R}^{\chictwo}_{\jpsi}$
related to the~photon reconstruction. 
The~corresponding systematic uncertainty
for the~efficiency ratio 
$\upvarepsilon^{\chicone}_{\chictwo}$
is found to be negligible.

The modelling of the~pion identification 
in the~simulation is improved by sampling 
the~corresponding distributions 
in the~\mbox{$\decay{\Dstarp}
{\left(\decay{\Dz}{\Km\pip}\right)\pip}$}
and \mbox{$\decay{\KS}{\pip\pim}$}~control 
channels~\cite{LHCb-DP-2012-003,
LHCb-DP-2018-001}. 
The~systematic uncertainty obtained
through this procedure arises from the~kernel 
shape used in the~estimation of the~probability
density distributions. An~alternative
%% combined 
response is estimated using 
%% an~alternative
a~different 
kernel estimation with a~changed shape and 
the~efficiency
models are regenerated~\cite{LHCb-PAPER-2020-025, Poluektov:2014rxa}.
The~difference between the~two estimates
for the~efficiency ratios is taken as the~systematic
uncertainty related to pion identification 
and is found to be 0.5\% and 0.6\% 
for the~ratio 
$\mathcal{R}^{\chictwo}_{\jpsi}$
and for the~efficiency ratio
$\upvarepsilon^{\chicone}_{\chictwo}$, respectively.

Large samples of~\mbox{$\decay{\Bp}{\jpsi\Kp}$} and 
\mbox{$\decay{\Bp}{\psitwos\Kp}$}~decays~\cite{LHCb-PAPER-2012-010}
are used to estimate the~systematic uncertainty 
related to the~trigger efficiency. 
A~conservative estimate of 
1.1\% for the~relative difference 
between data and simulation is taken
as the~corresponding systematic uncertainty~\cite{LHCb-PAPER-2012-010}.

Discrepancies in reconstructed quantities between 
the~simulated samples and  data,
%% The~imperfect data description in the~simulation 
due to effects other than those described above, 
are studied by varying the~{\sc{BDTG}}~selection criteria
for the~normalisation decay 
\mbox{$\decay{\Bc}{\jpsi\pip}$}
over its full range and repeating the~fit. 
The~observed maximal difference between 
the~efficiency estimated using 
data and simulation does not exceed 2.1\%.
This value is taken as a~corresponding 
systematic uncertainty
for the~ratio 
$\mathcal{R}^{\chictwo}_{\jpsi}$.
The~corresponding 
systematic uncertainty 
for the~efficiency ratio 
$\upvarepsilon^{\chicone}_{\chictwo}$
is considered to be negligible due to 
%% the~great  similarity of the~
the~very similar kinematics of the~corresponding
decay channels.

The~systematic uncertainty 
due to the~finite size of the~simulated samples,
used to calculate the~efficiency ratios
from Eqs.~\eqref{eq:effs}, 
is found to be  0.3\% 
and 0.4\%
for the~ratio 
%% $\mathcal{R}^{\BcTochictwopi}_{\BcTojpsipi}$
$\mathcal{R}^{\chictwo}_{\jpsi}$
and 
for the~efficiency ratio 
$\upvarepsilon^{\chicone}_{\chictwo}$, respectively.      
The~total systematic uncertainty is 
estimated as the~sum in quadrature of 
the~individual contributions.

For each choice of an~alternative fit model 
and for each pseudoexperiment
with random removal of multiple candidates, 
the~statistical significance for 
the~\mbox{$\decay{\Bc}{\chictwo\pip}$}~decay
is recalculated using Wilks’ theorem~\cite{Wilks:1938dza}. The~smallest significance of $7.6$~standard deviations 
is taken as the~overall significance 
of the~signal, including systematic uncertainties.

The~systematic uncertainty
for the~efficiency ratio 
$\upvarepsilon^{\chicone}_{\chictwo}$,
discussed above, is included 
in the~fit model via a~Gaussian constraint.
Therefore the~upper limit for
the~$\mathcal{R}^{\chicone}_{\chictwo}$~ratio 
from Eq.~\eqref{eq:f12_stat} already accounts for
the~corresponding systematic uncertainty. 
The~remaining uncertainties are related to 
the~fit model and the~exclusion of multiple candidates. 
To~account for these, 
the~upper limit is recalculated for 
each fit with the~alternative models
and for each pseudoexperiment
with random removal of multiple candidates
and the~maximal obtained value of %an~upper limit of
 \begin{equation*}
   \mathcal{R}^{\chicone}_{\chictwo} < 0.49~\text{at 90\%\,CL} 
 \end{equation*}
is conservatively taken as the~upper limit including 
systematic uncertainties.

\section{Results and conclusions}
\label{sec:Results}

The~decay modes \BcTochicjpi are studied
using proton\nobreakdash-proton collision data, 
corresponding to an~integrated luminosity of 
$9\invfb$,
collected with the~\lhcb detector at 
centre\nobreakdash-of\nobreakdash-mass energies of 7, 8, and 13\tev.
The~first observation of   
the~decay
\mbox{$\BcTochictwopi$}, with a~significance of 
$7.6$~standard deviations, is reported. 
Using the~\BcTojpsipi~decay as a~normalisation channel, and 
the~known value of the~branching fraction for 
the~\decay{\chictwo}{\jpsi\g} decay
%% $\BR_{\decay{\chictwo}{\jpsi\g}} = 
of~\mbox{$(19.0\pm0.5)\%$}~\cite{PDG2023},
the~ratio of the~branching fractions is measured to be
\begin{equation*}
\dfrac{\BR_{\BcTochictwopi}}{\BR_{\BcTojpsipi}}
%% \mathcal{R}^{\BcTochictwopi}_{\BcTojpsipi} 
=
%% 0.374 \pm 0.058 \pm 0.016 \pm 0.010 \,,   
   0.37  \pm 0.06  \pm 0.02 \pm 0.01 \,,
\end{equation*}
 where the~first uncertainty 
 is 
 statistical,
 %% from the~fit 
 the~second is systematic and 
 the~third is due to knowledge 
 of the~\decay{\chictwo}{\jpsi\g} 
 branching fraction~\cite{PDG2023}.
 The~measured value 
 is compatible with  
predictions
from Refs.~\cite{PhysRevD.74.074008}
 and~\cite{PhysRevD.73.054024}.

\begin{figure}[b]
	\setlength{\unitlength}{1mm}
	\centering
	\begin{picture}(150,95)
	%%
	%%\graphpaper[5](-10,-10)(170,140)
	%%
	\put(40,10){\includegraphics*[width=110mm,height=80mm]{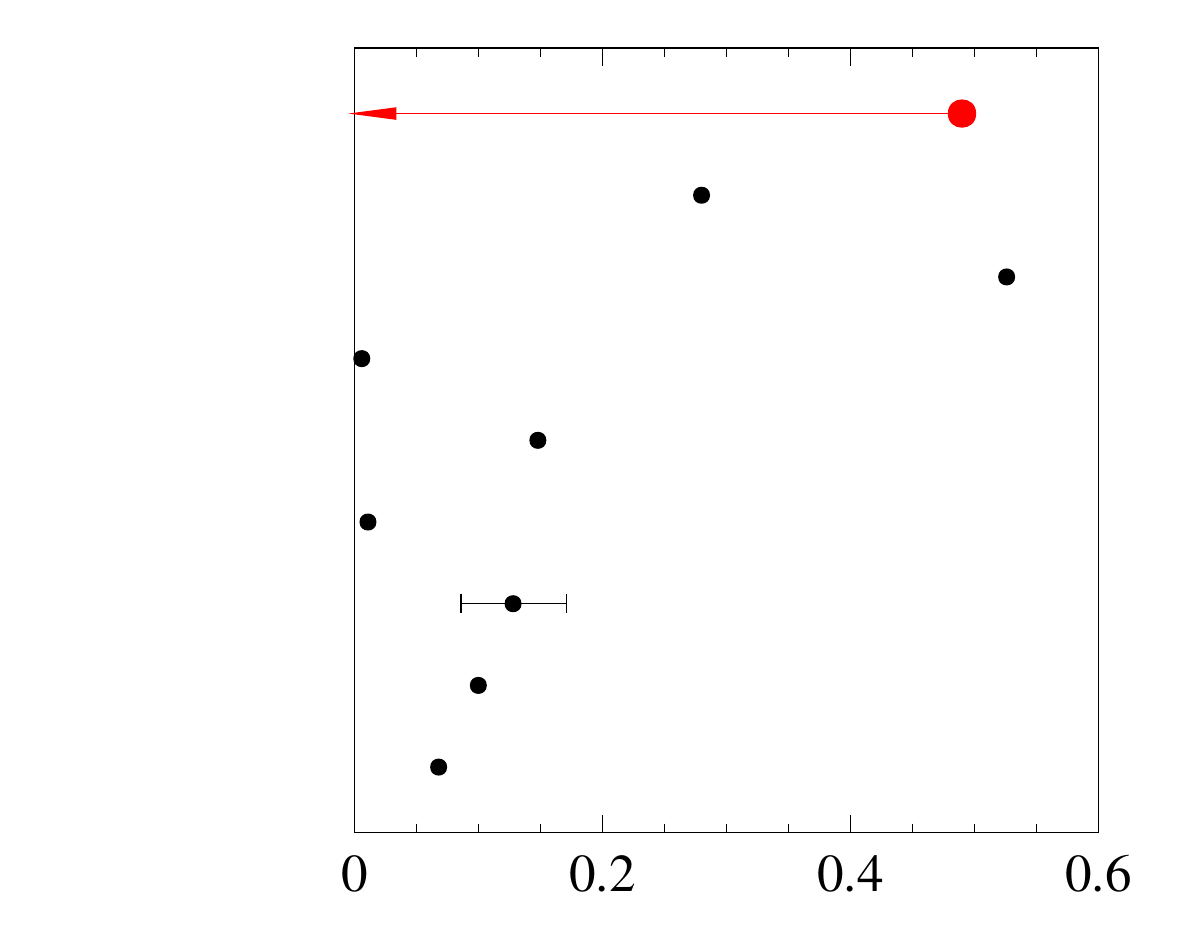}}
	%%\put(105,  3){\large $\mathcal{R}^{\chicone}_{\chictwo}$}
        \put(100,5){\large$\dfrac{\BR_{\BcTochiconepi}}{\BR_{\BcTochictwopi}}$}
   \put(5,48){\large\begin{Tabular}[1.1]{ll} 
    \textcolor{red}{LHCb\,2023 (90\%\,CL)}     &    \\
    C.-H.~Chang {\it{et al.}}                  &  \cite{PhysRevD.65.014017} \\
    D.~Ebert {\it{et al.}}                     &  \cite{PhysRevD.82.034019} \\ 
    E.~Hern\'andez {\it{et al.}}               &  \cite{PhysRevD.74.074008} \\
    M.~A.~Ivanov {\it{et al.}}                 &  \cite{PhysRevD.73.054024} \\ 
    V.~V.~Kiselev {\it{et al.}}                &  \cite{vvKiselev_2002}     \\ 
    Z.~Rui                                     &  \cite{PhysRevD.97.033001} \\ 
    Z.-h.~Wang {\it{et al.}}                   &  \cite{Wang_2012}          \\
    R.~Zhu                                     &  \cite{ZHU2018359}         \\\\
   \end{Tabular}}
	\end{picture}
    \caption{\small  
    Comparison of upper limit on the branching fraction 
    ratio between \BcTochiconepi and \BcTochictwopi 
    decays set in this analysis %% (red point with an~arrow)
    with theoretical predictions.}
\label{fig:uppers_cmp}
\end{figure}

No significant \mbox{$\decay{\Bc}{\chicone\pip}$}~signal 
is observed and 
%% The decay via intermediate \chicone state is found significantly suppressed.
the~upper limit on the~relative branching fraction between
the~\BcTochiconepi and \BcTochictwopi decays is found to be
\begin{equation*}
\dfrac{\BR_{\BcTochiconepi}}{\BR_{\BcTochictwopi}} 
%% \mathcal{R}^{\chicone}_{\chictwo}
   < 0.49~\text{at 90\%\,CL}\,. 
 \end{equation*}
 The~comparison of this upper limit with theoretical predictions, reported
 in Refs.~\cite{
 PhysRevD.65.014017,
 PhysRevD.82.034019,
 PhysRevD.74.074008,
 PhysRevD.73.054024,
 vvKiselev_2002,
 PhysRevD.97.033001,
 Wang_2012,
 ZHU2018359} is shown in Fig.~\ref{fig:uppers_cmp}. 
The~upper limit obtained is in 
agreement with all but one prediction
and disfavours
the~calculation   
from the~relativistic quark model~\cite{PhysRevD.82.034019}.  

%% file: acknowledgements.tex
\section*{Acknowledgements}
%
% These Acknowledgements valid from 3-May-2019
%
\noindent
We thank A.\,K.~Likhoded, X.~Liu and A.\,V.~Luchinsky
for the~interesting and stimulating discussions
on~the~physics of the~\Bc~mesons and their decays.
We~express our gratitude to our colleagues in the~CERN
accelerator departments for the excellent performance of the~LHC. 
We~thank the~technical and administrative staff at the~LHCb
institutes.
We~acknowledge support from CERN and from the~national agencies:
CAPES, CNPq, FAPERJ and FINEP\,(Brazil); 
MOST and NSFC\,(China); 
CNRS/IN2P3\,(France); 
BMBF, DFG and MPG\,(Germany); 
INFN\,(Italy); 
NWO\,(Netherlands); 
MNiSW and NCN\,(Poland); 
MCID/IFA\,(Romania); 
%MSHE (Russia); 
MICINN\,(Spain); 
SNSF and SER\,(Switzerland); 
NASU\,(Ukraine); 
STFC\,(United Kingdom); 
DOE NP and NSF\,(USA).
We~acknowledge the~computing resources that are provided by CERN, 
IN2P3\,(France), 
KIT and DESY\,(Germany), 
INFN\,(Italy), 
SURF\,(Netherlands),
PIC\,(Spain), 
GridPP\,(United Kingdom), 
%RRCKI and Yandex LLC (Russia), 
CSCS\,(Switzerland), 
IFIN\nobreakdash-HH\,(Romania), 
CBPF\,(Brazil),
and Polish WLCG\,(Poland).
We~are indebted to the~communities behind the~multiple open\nobreakdash-source
software packages on which we depend.
Individual groups or members have received support from
ARC and ARDC\,(Australia);
Key Research Program of Frontier Sciences of CAS, CAS PIFI, CAS CCEPP, 
Fundamental Research Funds for the~Central Universities, 
and Sci. \& Tech. Program of Guangzhou\,(China);
Minciencias\,(Colombia);
EPLANET, Marie Sk\l{}odowska\nobreakdash-Curie Actions, ERC and NextGenerationEU\,(European Union);
A*MIDEX, ANR, IPhU and Labex P2IO, and R\'{e}gion Auvergne\nobreakdash-Rh\^{o}ne\nobreakdash-Alpes\,(France);
%RFBR, RSF and Yandex LLC (Russia);
AvH Foundation\,(Germany);
ICSC\,(Italy); 
GVA, XuntaGal, GENCAT, Inditex, InTalent and Prog.~Atracci\'on Talento, CM\,(Spain);
SRC\,(Sweden);
the~Leverhulme Trust, the~Royal Society
 and UKRI\,(United Kingdom).

%% file: Authorship_LHCb-PAPER-2023-039.tex
% LHCb collaboration author list
% Data extracted on December 13th, 2023 at 5:00pm for paper reference LHCb-PAPER-2023-039
\centerline
{\large\bf LHCb collaboration}
\begin
{flushleft}
\small
R.~Aaij$^{35}$\lhcborcid{0000-0003-0533-1952},
A.S.W.~Abdelmotteleb$^{54}$\lhcborcid{0000-0001-7905-0542},
C.~Abellan~Beteta$^{48}$,
F.~Abudin{\'e}n$^{54}$\lhcborcid{0000-0002-6737-3528},
T.~Ackernley$^{58}$\lhcborcid{0000-0002-5951-3498},
A. A. ~Adefisoye$^{66}$\lhcborcid{0000-0003-2448-1550},
B.~Adeva$^{44}$\lhcborcid{0000-0001-9756-3712},
M.~Adinolfi$^{52}$\lhcborcid{0000-0002-1326-1264},
P.~Adlarson$^{79}$\lhcborcid{0000-0001-6280-3851},
C.~Agapopoulou$^{46}$\lhcborcid{0000-0002-2368-0147}
% ...

{\footnotesize \it

% ...
$^{35}$Nikhef National Institute for Subatomic Physics, Amsterdam, Netherlands\\
% ...
$^{44}$Instituto Galego de F{\'\i}sica de Altas Enerx{\'\i}as (IGFAE), Universidade de Santiago de Compostela, Santiago de Compostela, Spain\\
% ...
$^{46}$European Organization for Nuclear Research (CERN), Geneva, Switzerland\\
% ...
$^{48}$Physik-Institut, Universit{\"a}t Z{\"u}rich, Z{\"u}rich, Switzerland\\
% ...
$^{52}$H.H. Wills Physics Laboratory, University of Bristol, Bristol, United Kingdom\\
% ...
$^{54}$Department of Physics, University of Warwick, Coventry, United Kingdom\\
% ...
$^{57}$School of Physics and Astronomy, University of Glasgow, Glasgow, United Kingdom\\
$^{58}$Oliver Lodge Laboratory, University of Liverpool, Liverpool, United Kingdom\\
% ...
$^{66}$Syracuse University, Syracuse, NY, United States\\
% ...
$^{79}$Department of Physics and Astronomy, Uppsala University, Uppsala, Sweden, associated to $^{57}$\\
% ...
}
\end{flushleft}